\documentclass[12pt,preprint]{aastex}

\newcommand{\etal}{et~al.}
\newcommand{\CIVdblt}{{\rm C}\kern0.1em{\sc iv}~$\lambda\lambda 1548, 1550$}
\newcommand{\MgIIdblt}{{\rm Mg}\kern 0.1em{\sc ii}~$\lambda\lambda 2796, 2803$}
\newcommand{\SiIVdblt}{{\rm Si}\kern 0.1em{\sc iv}~$\lambda\lambda 1393, 1402$}
\newcommand{\NVdblt}{\hbox{{\rm N}\kern 0.1em{\sc v}~$\lambda\lambda 1239, 1243$}}
\newcommand{\OVIdblt}{{\rm O}\kern 0.1em{\sc vi}~$\lambda \lambda 1032, 1038$}
\newcommand{\CII}{\hbox{{\rm C}\kern 0.1em{\sc ii}}}
\newcommand{\CIII}{\hbox{{\rm C}\kern 0.1em{\sc iii}}}
\newcommand{\CIV}{\hbox{{\rm C}\kern 0.1em{\sc iv}}}
\newcommand{\HI}{\hbox{{\rm H}\kern 0.1em{\sc i}}}
\newcommand{\NaI}{\hbox{{\rm Na}\kern 0.1em{\sc i}}}
\newcommand{\HII}{\hbox{{\rm H}\kern 0.1em{\sc ii}}}
\newcommand{\HeI}{\hbox{{\rm He}\kern 0.1em{\sc i}}}
\newcommand{\HeII}{\hbox{{\rm He}\kern 0.1em{\sc ii}}}
\newcommand{\AlII}{\hbox{{\rm Al}\kern 0.1em{\sc ii}}}
\newcommand{\NII}{\hbox{{\rm N}\kern 0.1em{\sc ii}}}
\newcommand{\Lya}{\hbox{{\rm Ly}\kern 0.1em$\alpha$}}
\newcommand{\Lyb}{\hbox{{\rm Ly}\kern 0.1em$\beta$}}
\newcommand{\Lyg}{\hbox{{\rm Ly}\kern 0.1em$\gamma$}}
\newcommand{\Lyd}{\hbox{{\rm Ly}\kern 0.1em$\delta$}}
\newcommand{\Lye}{\hbox{{\rm Ly}\kern 0.1em$8$}}
\newcommand{\FeII}{\hbox{{\rm Fe}\kern 0.1em{\sc ii}}}
\newcommand{\MgI}{\hbox{{\rm Mg}\kern 0.1em{\sc i}}}
\newcommand{\MgII}{\hbox{{\rm Mg}\kern 0.1em{\sc ii}}}
\newcommand{\OVI}{\hbox{{\rm O}\kern 0.1em{\sc vi}}}
\newcommand{\OVII}{\hbox{{\rm O}\kern 0.1em{\sc vii}}}
\newcommand{\OVIII}{\hbox{{\rm O}\kern 0.1em{\sc viii}}}
\newcommand{\NV}{\hbox{{\rm N}\kern 0.1em{\sc v}}}
\newcommand{\SiII}{\hbox{{\rm Si}\kern 0.1em{\sc ii}}}
\newcommand{\SiIII}{\hbox{{\rm Si}\kern 0.1em{\sc iii}}}
\newcommand{\SiIV}{\hbox{{\rm Si}\kern 0.1em{\sc iv}}}
\newcommand{\kms}{\hbox{km~s$^{-1}$}}
\newcommand{\cmsq}{\hbox{cm$^{-2}$}}
\newcommand{\cc}{\hbox{cm$^{-3}$}}
\newcommand{\gtsim}{\protect\raisebox{-0.5ex}{$\:\stackrel{\textstyle >}{\sim}\:$}}

\begin{document}



\title{The Nature of Weak {\MgII} Absorbing Structures
\altaffilmark{1,2}}

\author{Nikola Milutinovi\'c\altaffilmark{3}, Jane R.
Rigby\altaffilmark{4}, Joseph R.  Masiero\altaffilmark{3,5},
Ryan S. Lynch\altaffilmark{3}, Chris Palma\altaffilmark{3},
and Jane C. Charlton\altaffilmark{3}}

\altaffiltext{1}{Based in part on observations obtained with the
NASA/ESA {\it Hubble Space Telescope}, which is operated by the STScI
for the Association of Universities for Research in Astronomy, Inc.,
under NASA contract NAS5--26555.}

\altaffiltext{2}{Based in part on observations obtained at the
W.~M. Keck Observatory, which is operated as a scientific partnership
among Caltech, the University of California, and NASA. The Observatory
was made possible by the generous financial support of the W.~M. Keck
Foundation.}

\altaffiltext{3}{Department of Astronomy and Astrophysics, The
Pennsylvania State University, University Park, PA 16802, {\it milni,
charlton, cpalma@astro.psu.edu}}

\altaffiltext{4}{Department of Astronomy, University of Arizona,
Tucson, AZ, 85721, {\it jrigby@as.arizona.edu}}

\altaffiltext{5}{Institute for Astronomy, University of Hawaii, 2680 Woodlawn
Drive, Honolulu, HI 96822, {\it masiero@ifa.hawaii.edu}}


\begin{abstract}
We consider geometries and possible physical models for weak low
ionization absorbers based on the relative incidence of low and high
ionization absorption systems.  To facilitate this, we present a
survey of weak low ionization absorption systems
($W_r(2796){\leq}0.3${\AA}) in 35 quasar spectra from the archive of
high resolution, ultra-violet {\it HST}/STIS data.  When possible, we
supplemented these spectra with Keck/HIRES and {\it HST}/FOS data that
cover more transitions over a larger range of wavelengths.  We found a
total of 16 metal-line systems, with low and/or high ionization
absorption detected.  It is known that the weak low ionization
absorbers (which we probe with {\MgIIdblt} or through the combination
of {\CII}$\lambda$1335 and {\SiII}$\lambda$1260) trace an abundant
population of metal-enriched regions (close to solar metallicity).
Generally, models show that these systems have a $\sim$10pc region of
higher density gas and a $\sim$1kpc region that represents a lower
density phase of higher ionization absorption.  The goal of our survey
was to compare absorption systems detected in low and/or high
ionization gas (the latter traced with {\CIVdblt} absorption).  We
find the following: 1. All but 1 of the 10 weak low ionization systems
have a related high ionization phase.  In 3 cases the high ionization
gas has only a single component, kinematically centered on the low
ionization absorption, and in the other 6 cases there are additional
high ionization components offset in velocity.  There is one system,
toward quasar 3C~273, without a high ionization cloud; 2. There are
just 6 systems with only a high ionization phase as compared to the 9
systems with both low and high ionization phases; 3. The high
ionization absorption in weak low ionization systems is, on average,
stronger than in systems with only high ionization absorption; 4. The
high ionization absorption in weak low ionization systems has similar
kinematic structure to that in high ionization only systems.  We find
that filamentary and sheetlike geometries are favored, due to the
relatively small observed cross-section of high ionization only
systems. Our statistical arguments suggest that, although low
ionization absorbers are not closely associated with luminous
galaxies, they arise in their immediate environments within the cosmic
web.

\end{abstract}



\section{Introduction}
\label{sec:intro}

Quasar absorption line spectroscopy is a powerful probe for the
investigation of gas in the cosmic web.  It offers a high sensitivity
that makes it possible to trace different gas phases over a large
range of redshifts, exploring both the intergalactic medium and a
variety of morphologies and evolutionary stages of galaxies.

The {\MgII} resonant doublet has been extensively used to trace metals
in galaxies at $0.3 < z < 2.2$, the redshift range for which {\MgII}
lies in the optical.  It is found that these low ionization systems,
with $W_r(2796) {\geq} 0.3$~{\AA}, produce Lyman limit breaks
\citep{archiveII}.  The gas kinematics, evident in high resolution
absorption profiles, are consistent with material in both the disks and
extended halos of the host galaxies \citep{kinmod,steidel02}.  Almost all
of these ``strong'' {\MgII} systems are found within $38h^{-1}$~kpc of
$>0.05L_K^*$ galaxies
\citep{bb91,bergeron92,lebrun93,sdp94,s95,3c336}.  Furthermore, both
observationally and theoretically, it is also expected that metal-rich
absorbing gas can be found out to distances of at least
$100h^{-1}$~kpc from luminous galaxies \citep{chen, cirkovic}.  This
gas can be detected through absorption in higher ionization species and/or
as weak low ionization absorption.

Weak low ionization absorbers, with $0.02 {\leq} W_r(2796) {\leq}
0.3$~{\AA}, constitute a significant fraction of the high {\HI} column
density regime ($15.8 < \log N({\HI}) < 16.8~{\rm cm}^{-2}$) of the
{\Lya} forest at $z\sim1$ \citep{weak2}. With $dN/dz = 1.74\pm0.10$ for
$0.4 < z < 1.4$, they are twice as likely to be found along the quasar line of sight
as strong low ionization absorbers \citep{weak1}.  The
majority of these systems arise in an optically thin, sub-Lyman limit
environment \citep{archiveII}.  Unlike strong absorbers, weak low
ionization absorbers typically are not associated with bright galaxies (with
$L{\geq}0.05L_{\star}$) to within a $50h^{-1}$~kpc impact parameter of
the quasar line of sight (\citet{weak1}; but see \citet{churchill} for exceptions).
This suggests that weak low
ionization systems are a physically different population from
strong absorbers, a result confirmed by the {\MgII} equivalent width
distribution measured from the Sloan Digital Sky Survey
\citep{nestor}.

Two thirds of weak low ionization absorbers have only a single narrow
component with a Doppler $b$ parameter of just a few {\kms}
\citep{weak1, weak2}.  Detection of {\FeII} in some of these systems
indicates a small ionization parameter and high density, which implies a
small size, $<10$~pc \citep{weak2}.  Based on photoionization models,
at least some single-cloud weak low ionization absorbers without detected
{\FeII} are constrained to have similarly small sizes
\citep{weak1634}.  The absence of a Lyman limit break and the strength
of the Lyman series lines provide constraints on the absorbers'
metallicities.  Surprisingly, the derived metallicities are almost
always greater than 10\% solar, and are often as high as the
solar value \citep{weak2,weak1634}.  Those absorbers with a large iron
to magnesium ratio must incorporate material from Type Ia supernovae,
in which case the metals would be generated ``in-situ''.  Their large
observed numbers along quasar lines of sight, in combination with the
small derived sizes, suggests that single-cloud low ionization
systems must be extremely abundant.  If they are assumed to have a
spherical geometry there would be over a million such structures per
bright galaxy.
Most of the single-cloud weak low ionization absorbers have an
associated higher ionization phase \footnote{\baselineskip =
0.5\baselineskip A physically distinct region (or regions) that has a
different density and/or temperature.}  that gives rise to {\CIV}
absorption \citep{weak2}.  Photoionization modeling of three systems
along the PG~$1634+706$ line of sight showed that low ionization
absorption arises in a relatively high density region ($\sim 0.01~{\rm
cm}^{-3}$) with a thickness of $0.1$--$100$~pc.  Higher ionization
absorption arises in one or more lower density regions ($\sim
10^{-3}~{\rm cm}^{-3}$), with sizes on the order of a kiloparsec
\citep{weak1634}.

The other third of the weak low ionization absorbers have multiple, rather
than single, low
ionization clouds.  Modeling of one of these systems, toward
PG~1634+704 led to a derived, low metallicity of
$\sim$3\% solar in the low ionization gas \citep{zonak}.  The
kinematics of the low and high ionization gas suggested an origin in a
pair of dwarf galaxies or in a symmetric wind from a starbursting
dwarf \citep{zonak}.  Also, \citet{rosenberg} have suggested that
local analogs to weak low ionization absorbers are related to winds
from starbursting dwarfs.  In the case of the weak low
ionization absorber at $z=0.0052$ toward 3C~273 \citep{tripp3c273}, there is a
post--starbursting dwarf galaxy, with a consistent redshift, at an
impact parameter of $50 h^{-1}$~kpc \citep{stocke}.  If winds from
dwarfs typically produce multiple-cloud, weak low ionization
absorption, then dwarf galaxies could be responsible for a large
fraction of that absorber population.  On the other hand, there is
evidence that some multiple-cloud weak low ionization absorbers are
produced by lines of sight through the outskirts of luminous galaxies
\citep{masiero, ding05}.  This subclass would simply be an extension
of the strong low ionization absorber population, which is directly
produced by luminous galaxies \citep{ding05}.

The single-cloud weak low ionization absorbers present a dilemma.
They are not closely related to luminous galaxies like strong
absorbers, yet they have high metallicities inconsistent with production in
the traditional gas phase of dwarfs.  \citet{weak2} suggested that
they might be produced in Population III, pre-galactic star clusters
or in supernovae remnants in very low luminosity galaxies, but these
are only examples from possible classes of scenarios.  In this paper,
we seek constraints on the nature of the physical structures that
produce the low and high ionization phase absorption in single--cloud,
weak low ionization absorption line systems.

{\it Our strategy is to compare the relative numbers and kinematics of
systems with detected absorption from both low and high ionization
phases to those of systems without absorption detected from one or the
other of those phases.}  For example, in principle, the low ionization
absorption could be produced in a small spherical region embedded in a
much larger spherical structure that produces {\CIV} absorption.  In
this case, we would expect that most systems have only high ionization
absorption.  We use the archive of
{\it Hubble Space Telescope} ({\it HST})/ Space Telescope Imaging
Spectrograph (STIS) Echelle spectra ($R=30,000$ or $45,000$)
\citep{kimble} in order to make these comparisons, considering systems
with coverage of {\CIV} and at least some low ionization transitions
({\SiII}$\lambda$1260, {\CII}$\lambda$1335, and/or {\MgIIdblt}).


In \S~\ref{sec:data}, we briefly describe the characteristics of
our sample, and the search method.
In \S~\ref{sec:systems_found} we describe the specific
systems that we found.  In \S~\ref{sec:rez}, we present
statistical results and Voigt profile fits. In the discussion of the
paper, \S~\ref{sec:disc} and \S~\ref{sec:disc2}, we apply our
results to simple thought experiments in order to constrain possible
geometries of the systems, and finally to connect these models to
physical interpretations.

\section{Data and Method}
\label{sec:data}

To facilitate our study, we searched for absorption systems at $0<z<1$
for which high resolution coverage of both low and high
ionization transitions is available.  The {\it HST}/STIS
\citep{kimble} Echelle archive provides
an excellent database for this study. When possible, we supplemented
these data with Keck I/High Resolution Spectrograph (HIRES) \citep{vogt94} spectra
in order to obtain coverage of {\MgII} and {\FeII}.
For a few systems, the high resolution data were supplemented by data obtained
with the {\it HST}/Faint Object Spectrograph (FOS), which provided additional
wavelength coverage at lower resolution.

\subsection{STIS Archive}
\label{sec:stis}

We searched $35$ Echelle spectra of quasars from the {\it HST}/STIS
archive that were available before May 2004
(Tables~\ref{tab:tab1} and \ref{tab:tab2}).  These represented almost all of the
available Echelle datasets. We eliminated several spectra due to
poor $S/N$ ($<5$ per pixel), since these
are not suitable for a study of weak low ionization
absorption.  All the spectra were obtained with the medium resolution
gratings, E140M, with $R=45,800$, and E230M, with $R=30,000$.  These
resolutions correspond to $6.5$~{\kms} and $10$~{\kms}, respectively.
$S/N = 5$ corresponds to an observed equivalent width limit between $\sim
0.02$ and $0.03$~{\AA} in the E140M spectra, and between $\sim 0.06$ and
$0.1$~{\AA} in the E230M spectra.  For data reduction, we used the standard
STIS pipeline \citep{brown}.  Separate exposures and overlapping orders
were combined using standard procedures described in \citet{anand}.


\subsection{Keck/HIRES}
\label{sec:hires}

Since {\it HST}/STIS does not cover of {\MgIIdblt} for
the observed redshift interval, we supplement the {\it HST}/STIS data
with Keck I/HIRES data for PG~$1634+706$, PG~$1206+459$,
PG~$1248+401$, PG~$1241+176$, PG~$0117+210$, CSO~$873$, and
PG~$0454-220$ \citep{weak1,cv01}.  For many systems for which {\MgII}
was not covered, {\SiII}$\lambda$1260 and {\CII}$\lambda$1335 were used to represent
the strength of low ionization absorption. Since {\MgII} is the most
common tracer of low ionization systems at $z>0.4$, for calibration
purposes we used the Keck/HIRES dataset in conjunction with {\it HST}/STIS
to find systems with all three transitions, {\SiII}, {\CII}, and {\MgII},
detected (see \S~\ref{sec:search_method}).

The optical spectra were obtained with the HIRES spectrograph \citep{vogt94}
on the Keck I telescope. The resolution is $R=45,000$, corresponding
to $6.6$~{\kms}. These spectra have signal to noise
ratios ranging from $15$ to $30$ (per pixel) in the continuum near {\MgII}.

\subsection{FOS}
\label{sec:fos}

The {\it HST}/FOS data provided wavelength coverage of the {\Lya}
transition for the few systems for which this transition was not
covered by higher resolution {\it HST}/STIS observations.  The {\it
HST}/FOS spectra were obtained with the G130 or G190 gratings, which
had resolution $R=1,300$, corresponding to $\sim230~{\kms}$.

\subsection{Search Method}
\label{sec:search_method}

The goal of our survey was to detect systems at $0 < z < 1$ which have
high resolution wavelength coverage of both {\CIV} and representative
low ionization transitions ({\MgIIdblt}, and/or {\SiII}$\lambda$1260 and
{\CII}$\lambda$1335).

We first searched for the {\CIVdblt} doublet in the $35$ quasar spectra from
the STIS/Echelle archive. We compiled a list of $5~\sigma$ detections
using the method described in \citet{schneider93}. We assumed each
detection to be a candidate for the {\CIV}$\lambda$1548 line,
the stronger transition of the {\CIV} doublet. Then we
searched for absorption at the corresponding {\CIV}$\lambda1550$ line
position. A doublet candidate was designated if the significance of
the detection at that position was $>3~\sigma$, scaled down from
$5~\sigma$ by the expected {\CIV} doublet ratio.  For each of the doublet
candidates we searched the {\it HST}/STIS spectrum for the {\Lya}
transition at the same redshift. Detection of {\Lya} absorption over
the full velocity range of the {\CIV} detection was one criterion used
to assess the reality of the doublet. If {\Lya} was not covered by
the {\it HST}/STIS spectrum, we considered whether it was detected in
any available spectra from {\it HST}/FOS.  For each doublet candidate,
we also visually inspected the alignment and the profile shapes of the
doublet lines.  If the lines closely correspond to one another, and
there is no contradictory indication from {\Lya}, we consider the
system to be ``real''.  In order to be included in our sample we also required
that the $S/N$ and coverage of available spectra allowed for detection
of low ionization absorption tracers.

We measured the equivalent widths or equivalent width limits at
the expected positions of the low ionization features in all of the
``real'' systems.  For some systems, {\MgIIdblt} was covered in
our Keck/HIRES data.  However, when those data were not available,
since {\MgIIdblt} is not covered in the {\it HST}/STIS
spectra for absorbers at $z>0.1$, we used the {\SiII}$\lambda$1260 and
{\CII}$\lambda$1335 transitions as tracers of the low ionization phase.
We formed a calibration sample, including
all systems for which we have coverage of {\MgIIdblt} and either {\CII}$\lambda$1335
or {\SiII}$\lambda$1260.  To our survey systems, we added five additional
systems that failed to satisfy our survey requirements (e.g., have no coverage
of {\CIV}), but have the necessary coverage for this calibration.
These additional systems are found toward PG~0117+210 at $z=0.7290$, $1.3250$, and
$1.3430$, toward PG~1206+459 at $z=0.9343$, and toward
PG1634+706 at $z=1.0400$. 

The equivalent widths of {\MgII}$\lambda$2796, {\SiII}$\lambda$1260,
and {\CII}$\lambda$1335 for a given system are highly correlated.  The
probability that the distribution of {\MgII}$\lambda$2796 relative to
{\SiII}$\lambda$1260 is drawn from a random distribution is only
$0.17$\% by the Student's t-Test, with a correlation coefficient of
$0.88$.  Similarly, {\MgII}$\lambda$2796 is closely correlated
with {\CII}$\lambda$1335, with a correlation coefficient of
$0.87$, and with only a $0.21$\% chance of being drawn from a
random distribution.  The mean ratio $W_r({\CII}\lambda{\rm
1335})/W_r({\MgII}\lambda{\rm 2796}) = 1.07\pm0.49$ for the $9$
relevant systems in our calibration sample.  Similarly, the ratio,
$W_r({\SiII}\lambda{\rm 1260})/W_r({\MgII}\lambda{\rm 2796}) =
0.52\pm0.20$, determined from the same number of systems.

Previous surveys for weak {\MgII} absorption \citep{weak1,anand} applied
a $5\sigma$ rest frame equivalent width detection threshold of $0.02${\AA}.
Thus if we are sensitive to {\SiII}$\lambda$1260 to a $3~\sigma$ limit of
$W_r(1260)\gtsim 0.01${\AA} and to {\CII}$\lambda$1335 to a $3~\sigma$ limit
of $W_r(1335)\gtsim 0.02${\AA}, according to the mean ratios above.
Most of the relevant spectral coverage allows for detection of features
significantly weaker than these limits, therefore, even with the observed variation
in these ratios, we are sensitive to all systems equivalent to weak
{\MgII} absorbers.  Thus we can obtain an accurate indication of the fraction
of systems with detected {\CIV} that are equivalent to weak {\MgII} absorbers.

We also searched for systems which could be detected through low
ionization absorption, but for which {\CIV} absorption was not
detected.  We again used {\SiII}$\lambda$1260 and {\CII}$\lambda$1335
as tracers of weak {\MgII} systems, as in \citet{anand}.

\section{Systems}
\label{sec:systems_found}

We found 9 systems with both high and low ionization transitions
detected, 6 systems with high ionization only, and 1 with low
ionization only.  We note that, as described in \S~\ref{sec:search_method},
the detection limits for low ionization transitions associated with the 6
high ionization only systems were well below the traditional
$0.02$~{\AA} detection limit for weak {\MgII} absorbers
\citep{weak1,anand}.
For convenient reference, in the
headings for each system we give a simple classification.  ``SC''
and ``MC'' denote single and multiple-cloud systems, respectively.
The words ``low'' and ``high'' refer to the low ({\SiII}, {\CII},
and {\MgII}) and high ionization ({\CIV}) transitions.  If the word
``low'' or ``high'' is omitted, that means that these transitions
were covered, but were not detected in that system.

\subsection{System Information}

\subsubsection{System 1 -- 3c273, $z_{sys}=0.0052$, SC low}

This system, presented in Figure \ref{fig:f1}, is a single-cloud
weak absorber, with only low ionization absorption detected. The low
ionization absorption is detected as a single component in both
{\SiII}$\lambda$1260 and {\CII}$\lambda$1335.  The normalized
{\SiIII}$\lambda$1207 profile
was produced by ``fitting out'' Galactic {\Lya}.  {\CIVdblt} is not
detected to a $3~\sigma$ limit of $W_r(1548)<0.0034$~{\AA}. The
{\Lya}, which was superimposed on Galactic {\Lya}, is centered on
the low ionization absorption.  For convenience of display, we
divided by a model for the Galactic absorption in order to show the
shape of the {\Lya} profile for this system.  This system was
reported by \citet{tripp3c273}, and later discussed by
\citet{stocke} who reported a post-starburst galaxy at an impact
parameter of $50 h^{-1}$~kpc.

\subsubsection{System 2 -- RXJ~1230.8+0115, $z_{sys}=0.0057$, MC low + SC high}

This system, presented in Figure \ref{fig:f2}, is a multiple-cloud
weak low ionization absorber, detected in {\SiII} and {\CII}.  There are
two components apparent in both {\SiII}$\lambda$1260 and
{\CII}$\lambda$1335, which are also detected in
{\SiIV}$\lambda$1393 and {\SiIV}$\lambda$1402.  The region in
which {\SiIII}$\lambda$1207 would be observed is badly blended with
Galactic {\Lya}. {\CIV}$\lambda$1548 is detected in the stronger,
redward component, but not in the blueward component for which we
derive a limit $W_r(1548)<0.02$~{\AA}.  The {\Lya} line is strong
and roughly centered on the two components.  This system was
published previously and discussed by \citet{rosenberg}, who also
published FUSE data covering the Lyman series lines.

\subsubsection{System 3 --  PG 1211+143, $z_{sys}=0.0512$, SC low + MC high}

This system is a single-cloud weak low ionization absorber with
multiple-component high ionization absorption. It is shown in Figure
\ref{fig:f3}. Low ionization absorption is detected in {\SiII} and
{\CII}.  The blueward portion of the {\CII}$\lambda$1335 profile is
blended with Galactic {\SiIVdblt}.  The {\CIV} profile has three
components, with the strongest centered on the single-cloud low
ionization feature.  The {\SiIV} doublet lines show two components.
The stronger one is centered on the low ionization absorption, and
the weaker, blueward component is centered on the coincident {\CIV}
cloud. {\SiIII}$\lambda$1207 is detected, but it is blended with a {\Lya} line at
$z=0.0435$.  The broad {\Lya} line is asymmetric relative to both
the low and high ionization absorption, extending further to the
red. This system was discussed by \citet{stocke}.

\subsubsection{System 4 -- PHL 1811, $z_{sys}=0.0809$, SC low + SC high}

This system, presented in Figure \ref{fig:f4}, is a single-cloud
weak low ionization system, with a single component high ionization
profile.  Low ionization absorption is detected in
{\SiII}$\lambda$1190, {\SiII}$\lambda$1193, {\SiII}$\lambda$1260,
and {\SiII}$\lambda$1527, as well as in {\CII}$\lambda$1335. The
blueward part of the {\SiII}$\lambda$1260 line is blended with a
{\Lya} line at $z = 0.1205$.
The intermediate ionization transition, {\SiIII}$\lambda$1207, is
blended to the red with the Galactic {\SiII}$\lambda$1304 line but
it can be clearly separated. High ionization absorption is detected
in {\SiIVdblt} and {\CIVdblt}. The quality of the spectrum at the
position of {\CIVdblt} is just higher than our survey threshold of
$S/N = 5$.  Because of this, it was difficult to assess whether
additional components were detected in {\CIV}, particularly one at
$\sim-50$~{\kms}.  The {\Lya} profile is broad and asymmetric, and
is not centered on the profiles of the metal-line transitions.  This
system was published and studied by \citet{jenkins03}.

\subsubsection{System 5 -- PG 1241+174, $z_{sys}=0.5584$, MC low
+ MC high}

This multiple-cloud weak low ionization system is presented in
Figure \ref{fig:f5}.  The {\MgII} doublet is detected in the
Keck/HIRES spectrum. {\CII}$\lambda$1335 and {\SiII}$\lambda$1260
are not covered in the {\it HST}/STIS spectrum.  The weaker
{\SiII}$\lambda$1527 and {\AlII}$\lambda$1671 low ionization
transitions are covered in the {\it HST}/STIS spectrum, but are not
detected at $5~{\sigma}$. {\CIVdblt} is detected in the {\it
HST}/STIS spectrum.  Both the high and low ionization absorption
profiles require multiple component fits.  The {\CIV} profiles are
roughly centered on the low ionization profiles.  {\Lya} is not
covered by either spectrum.  This system is reported in
\citet{weak1} and modeled by \citet{ding05}.

\subsubsection{System 6 -- PG 1634+706, $z_{sys}=0.6534$, SC low + MC high}

This is a single-cloud weak low ionization absorption system with
multiple components detected in {\CIV}, as well as in {\SiIV}. The
system is shown in Figure \ref{fig:f6}. {\SiII}$\lambda$1260 and
{\CII}$\lambda$1335 are detected in the {\it HST}/STIS data. 
{\MgIIdblt} is detected in the Keck/HIRES spectrum.  The
strongest high ionization feature is centered on the low ionization
cloud. 
{\SiIII}$\lambda$1207 is detected, but is blended to the blue,
possibly with {\Lya}.  {\Lya} is broad, and roughly centered on the
high ionization absorption. Previous modeling by \citet{weak1634}
shows separate low and high ionization phases.

\subsubsection{System 7 -- PG 1634+706, $z_{sys}=0.8181$, SC low + SC high}

This system is a single-cloud weak low ionization absorber with
detected {\MgIIdblt}, {\SiII}$\lambda$1260, and
{\CII}$\lambda$1335 (see Figure \ref{fig:f7}). The {\MgIIdblt} were
covered in the Keck/HIRES spectrum. Coverage of other transitions
was provided through the {\it HST}/STIS dataset. The {\Lya} and
{\CIV} are centered on the {\MgII} cloud and do not have offset
components.  The {\SiIII}$\lambda$1207 transition is blended,
probably with {\Lya} at $z=0.80412$.  This system was modeled and
discussed by \citet{weak2} and \citet{weak1634}.  Although the low
and high ionization absorption are centered at the same velocity,
they must arise in two different phases (i.e., regions with
different ionization parameters) \citep{weak1634}.

\subsubsection{System 8 -- PG 1248+401, $z_{sys}=0.8548$, MC low + MC high}

This system, shown in Figure \ref{fig:f8}, was detected in both {\it
HST}/STIS and Keck/HIRES spectra.  It shows complex multi-cloud
structure in both the low and high ionization transitions. Higher
ionization features are centered on the low ionization absorption,
but do not trace it in strength. The {\CII}$\lambda$1335 profile
is contaminated by an unknown feature at $\sim$160~{\kms}, and there
is the possibility that other redward components are affected as
well.  The redward components of
{\SiIV}$\lambda$1393 are heavily blended with Galactic
{\FeII}$\lambda$2587. {\NVdblt} may also be detected.
The weaker line in that doublet is blended to the blue with {\Lya}
at $z=0.8951$. {\Lya} was not covered by the {\it HST}/STIS
spectrum.  This system was modeled and discussed by \citet{ding05}.

\subsubsection{System 9 -- PG 1241+174, $z_{sys}=0.8954$, SC low + SC high}

This system, presented in Figure \ref{fig:f9}, is a single-cloud
weak, low ionization system. The {\MgII} resonant doublet is
detected in a Keck/HIRES spectrum. The {\CIV} and {\SiIV} doublets
are detected in {\it HST}/STIS data, though {\SiIV}$\lambda$1402 is
affected by an unidentified blend, possibly {\Lya}.  These high
ionization profiles can be fit with single components, centered on
the {\MgII}. {\CII}$\lambda$1335 is not detected in a noisy region
of the {\it HST}/STIS spectrum.  There is a $5~\sigma$ detection
at the expected position of {\SiII}$\lambda$1260, however, based on
photoionization models \citep{ding05}, the {\SiII}$\lambda$1260
seems too strong relative to {\MgII}$\lambda$2796 (see
Table~\ref{tab:eqwidths} for measurements and limits).  The {\it HST}/STIS
spectrum alone was not adequate to assess the classification of this
system, but since we had access to Keck/HIRES data we were able to
include it in our sample. Modeling of the system was complicated by
an apparent offset ($\sim 7$~{\kms}) between {\MgII} and {\CIV}.  If
the offset is neglected, the {\CIV} could arise in the same phase
with the {\MgII}, but two-phase models are also consistent with the
data \citep{ding05}.

\subsubsection{System 10 -- PG 1634+706, $z_{sys}=0.9056$, SC low + MC high}

This is a single-cloud weak low ionization system (see Figure
\ref{fig:f10}). {\MgIIdblt} is detected in Keck/HIRES data.
{\SiII}$\lambda$1260, {\CII}$\lambda$1335, and
{\SiIII}$\lambda$1207 are detected in the {\it HST}/STIS spectrum.
The {\SiII}$\lambda$1260 line is slightly blended to the blue
with {\SiIII}$\lambda$1207 at $z=0.9902$, but can be clearly
separated.  {\CIVdblt} is detected at the velocity of the low
ionization absorption, but the profiles are asymmetric, with an
offset to the red.  The {\OVIdblt} doublet is covered and detected.
Detailed discussions of this system appear in \citet{weak2} and
\citet{weak1634}. Modeling showed that even the {\CIV} that was
centered on the low ionization absorption must come from a separate,
higher ionization phase \citep{weak1634}.


\subsubsection{System 11 -- HS0624+6907, $z_{sys}=0.0635$, MC high}

This system, presented in Figure \ref{fig:f11}, is a high ionization only
(high only) system. The {\CIV} doublet is detected in three components.  The high
ionization absorption is confirmed for both components of
{\SiIVdblt}, however, the detection of the weaker, redward component
of {\SiIV}~$1402$ is just above the $5~\sigma$ limit. Absorption is
also detected in {\SiIII}$\lambda$1207 for both components. The
{\Lya} profile extends over the same velocities as the three high
ionization components.


\subsubsection{System 12 -- PG1211+143, $z_{sys}=0.0644$, MC high}

This is a high only system (see Figure
\ref{fig:f12}). It has multiple, but weak, {\CIV} components. {\SiIV}
is not detected, however, there may be a detection of {\SiIII}$\lambda$1207.
{\Lya} is saturated and it covers the same velocities as the {\CIV}
absorption.

\subsubsection{System 13 -- HS0624+6907, $z_{sys}=0.0757$, SC high}

The system, shown in Figure \ref{fig:f13}, is characterized by
single-cloud high ionization absorption, which is detected as weak
{\CIVdblt}.  Both of the strong low ionization indicators,
{\CII}$\lambda$1335 and {\SiII}$\lambda$1260 are covered, but
absorption is not detected. Lines of the {\CIV} doublet are weak, with
the minimum aligned in velocity with the broader {\Lya}.

\subsubsection{System 14 -- PG1206+459, $z_{sys}=0.7338$, MC high}

This system, presented in Figure \ref{fig:f14}, is a multiple-cloud
high only system. The {\CIVdblt} doublet is detected in two
components.  The reddest component is the strongest.  The bluest
component is somewhat disputable.  Its {\CIV}$\lambda$1550
transition is detected at the $3~\sigma$ level, but it seems too weak
compared to the matching component in {\CIV}$\lambda$1548.  The
{\CIV}$\lambda$1548 line is blended with the {\SiIV}$\lambda$1393
absorption from a system at $z=0.9254$ \citep{ding1206}. There is
also an unidentified blend redward of the strongest component.  The
{\CIV}$\lambda$1550 line is also blended, to the blue, with the
{\SiIV}$\lambda$1393 line of an absorption system at $z=0.9276$,
but these two lines can be cleanly separated.  We use
{\CIV}$\lambda$1550 to determine the two component fit 
listed in Table~\ref{tab:vpfits}.
{\CII}$\lambda$1335 is covered, but there is no absorption detected
at that wavelength.  {\MgIIdblt} is also not detected in the
Keck/HIRES spectrum.  {\Lya} is not covered in the {\it HST}/STIS
spectrum.

\subsubsection{System 15 -- PG1248+401, $z_{sys}=0.7011$, SC high}

The system shown on  \ref{fig:f15} is a single-cloud high only
absorption system. The only detected absorption is in {\CIVdblt}.
The {\SiIVdblt} doublet is covered, but there is no detected
absorption.  The wavelengths at the expected position of
{\SiIV}$\lambda$1402 are contaminated by {\Lya} at $z=0.9635$.
{\Lya} at $z_{sys}$ is not covered in the HST/STIS spectrum.

\subsubsection{System 16 -- PG1630+377, $z_{sys}=0.9143$, MC high}
\label{sec:sys16}

This system, shown in Figure \ref{fig:f16}, has a {\CIV} doublet that
can be adequately fit with three components.  Because the {\CIV}$\lambda$1548
profile is heavily blended to the blue, our fit to the bluer part of the profile
was determined only using {\CIV}$\lambda$1551.
The expected position of {\CII}$\lambda$1335 is affected by an unidentified
blend, but there is a feature detected at $3~{\sigma}$ at the
wavelength of the {\SiII}$\lambda$1260 line, which is centered
on the strongest {\CIV} component.  We classify this system as high only,
however this classification is uncertain, with a
low ionization phase being possible.
There are also detections of {\SiIII}$\lambda$1207 and of the
{\SiIV} and {\NV} doublets.  The {\NV}$\lambda$1243 line is blended
with {\Lya} at $z=0.9573$.  The {\Lya} profile is not symmetric
but is extended to the red of the strongest {\CIV} component,
presumably due to a contribution from the cloud that produced the
redward {\CIV} component.

\subsection{Systems That Failed To Pass The Survey Requirements}

For some systems in our dataset we did not have the required coverage
of {\CIVdblt} or of low ionization transitions. In that case we could
sometimes supplement our dataset with the {\it HST}/FOS Archive.
However, since the resolution
of FOS is insufficient for our study of {\CIV} kinematics, we exclude them
from our main survey.  For some systems, {\CIV} was covered in a
{\it HST}/STIS dataset, however that region had $S/N$ less than
our requirement of $5$ per pixel.
We excluded the following systems from our survey:
$z=0.1385$ toward PG 1116+215, $z=0.7390$ toward PKS 0232-04,
$z=0.8313$ toward HS 0810+2554, $z=0.5648$ toward
PG 1248+401, $z=0.7290$, $1.3250$, and $1.3430$ toward PG~0117+210,
$z=0.9343$ toward PG~1206+459, and $z=1.0400$ toward
PG1634+706.

\section{Results}
\label{sec:rez}

We classify systems based upon whether they have detected low
and/or high ionization transitions.  Specifically, we define ``weak
low ionization + high ionization'' (low + high), ``low ionization only''
(low only), and ``high ionization only'' (high only) systems.  Here
we consider the relative numbers of these three types of systems
in our unbiased survey.

Nine of the systems found in our survey have both high and low
ionization transitions detected (six have single-cloud low
ionization components and three have multiple), six systems are high
ionization only, and one system has only low ionization absorption
detected.  Table~\ref{tab:eqwidths} lists the equivalent widths of
key transitions for these three categories of systems.
Four of the six single-cloud low + high systems
have multiple clouds in their high ionization phases. Four of the
six high ionization only systems have multiple-component clouds.


\subsection{Comparison of {\CIV} Profiles -- Kinematics of {\CIV}}

In Figure~\ref{fig:civkinplt}, we compare the kinematic profiles of
{\CIV} between the different classes of systems.  All but one system,
which only has low ionization absorption detected (the $z=0.0053$ system
toward 3C~273), have high ionization absorption at the same velocity
as their low ionization absorption.  In some cases, there is only one
component detected in {\CIV}, but in most cases there are additional
{\CIV} components offset by $5$-$150$~{\kms}.  The {\CIV} component
that is centered on the low ionization absorption is always stronger
than the offset {\CIV} component/s.  In one case (the $z=0.0057$ system
toward RX~J1230.8+0115) one of the two low ionization components does not
have detected {\CIV} absorption.  The high only category of absorbers
also contains both single and multiple component {\CIV} profiles.

The kinematic spreads of the {\CIV} in high only systems are similar to those of
the {\CIV} of systems in the low + high ionization category.  There is, however,
an important difference between the {\CIV} profiles of high only and
low + high ionization systems.  To illustrate this, we performed
Voigt profile fits to the {\CIV} profiles for these systems and we
present the results in Table~\ref{tab:vpfits}.
Figure~\ref{fig:logn} shows the {\CIV} column densities for the
individual Voigt profile components in each system, allowing
comparison between the {\CIV} in low + high and high only systems.
Although there is some overlap, high only systems tend to have
component {\CIV} column densities less than those of low + high
ionization systems.  A Kolmogorov-Smirnov (KS) test between the
two samples yields only a $2.7$\% chance that they are drawn from
the same distribution.

The strongest {\CIV} components in the high only systems generally fall in the
same column density range as the offset {\CIV} components of low + high
systems.  A Kolmogorov-Smirnov (KS) test between
the distributions of maximum {\CIV} component column densities for
these two samples yields a $14$\% chance that they are drawn
from the same distribution, which is not a very significant difference.
However, the samples are small, and if System 16 is considered to be
a low + high system (possible because of the $3~\sigma$ detection
of {\SiII}$\lambda$1260, as discussed in \S~\ref{sec:sys16}), there is
only a $4.1$\% chance that they are drawn from the same distribution.

\subsection{Comparison of Lya Profiles}

\begin{figure*}
\figurenum{1} \plotone{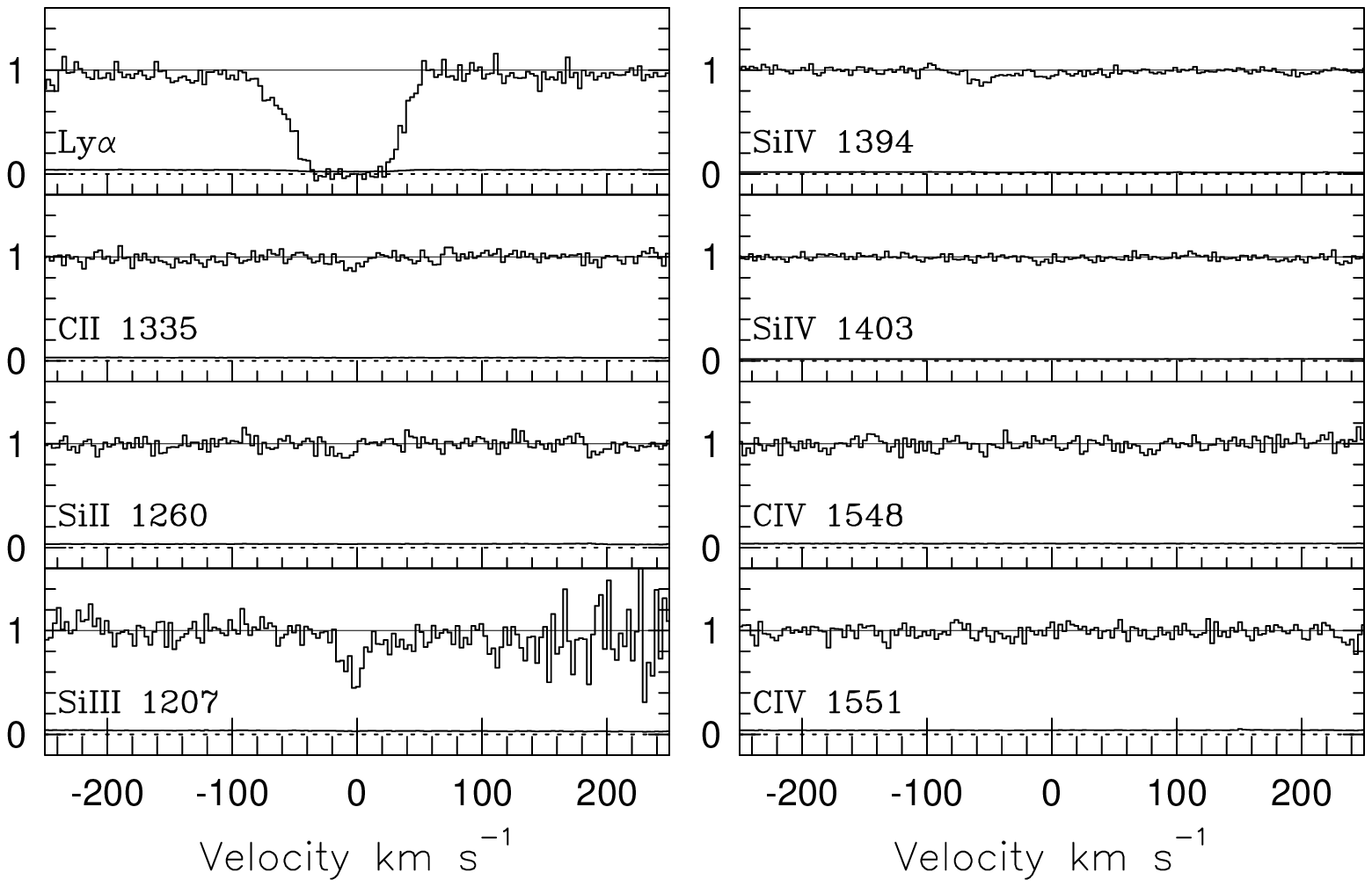} \protect \caption{ Absorption
profiles for key transitions, displayed as normalized flux vs.
velocity, for the $z =0.0052$ system toward the quasar 3C 273.  All
data are from an {\it HST}/STIS spectrum.} \label{fig:f1}
\end{figure*}

\begin{figure*}
\figurenum{2} \plotone{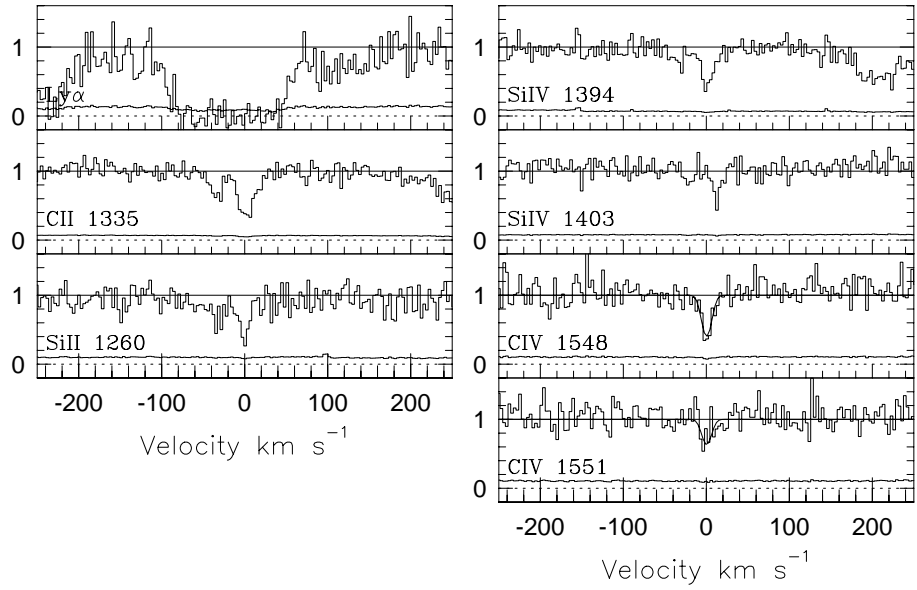} \protect \caption{Same as
Figure~\ref{fig:f1}, for the $z$ = 0.0057 system toward RXJ
1230.8+0115 quasar spectrum. Results of our Voigt profile fit to {\CIVdblt}
are superimposed as a solid curve on the data.} \label{fig:f2}
\end{figure*}

\begin{figure*}
\figurenum{3} \plotone{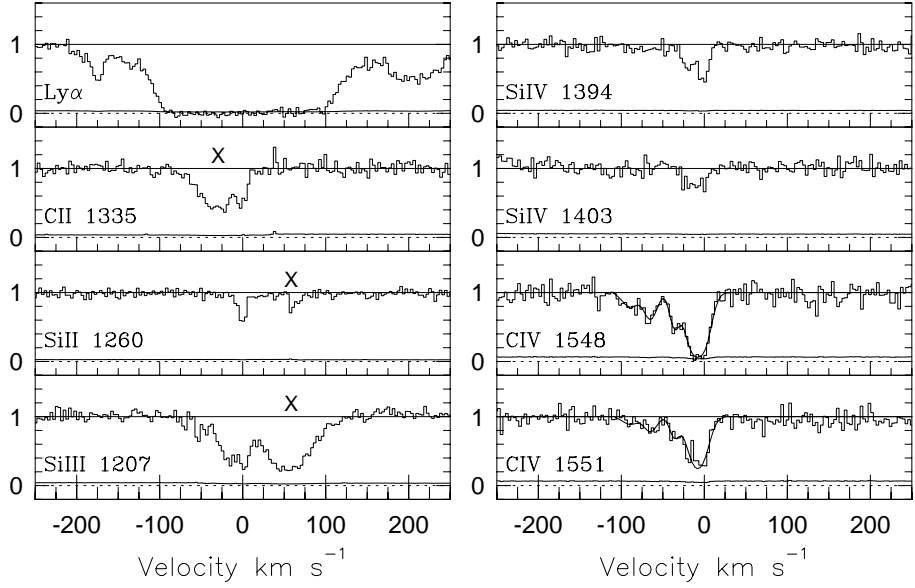} \vglue -1in \protect \caption{Same as
Figure~\ref{fig:f2}, for the $z$ = 0.0512 system toward PG 1211+143.
The {\CII}$\lambda$1335 transition is blended to the blue with
Galactic {\SiIV}. {\SiIII} is also detected, but it is blended with
a {\Lya} line at $z=0.0435$.  Blends are marked with the symbol ``X''.} \label{fig:f3}
\end{figure*}

\begin{figure*}
\figurenum{4}
\plotone{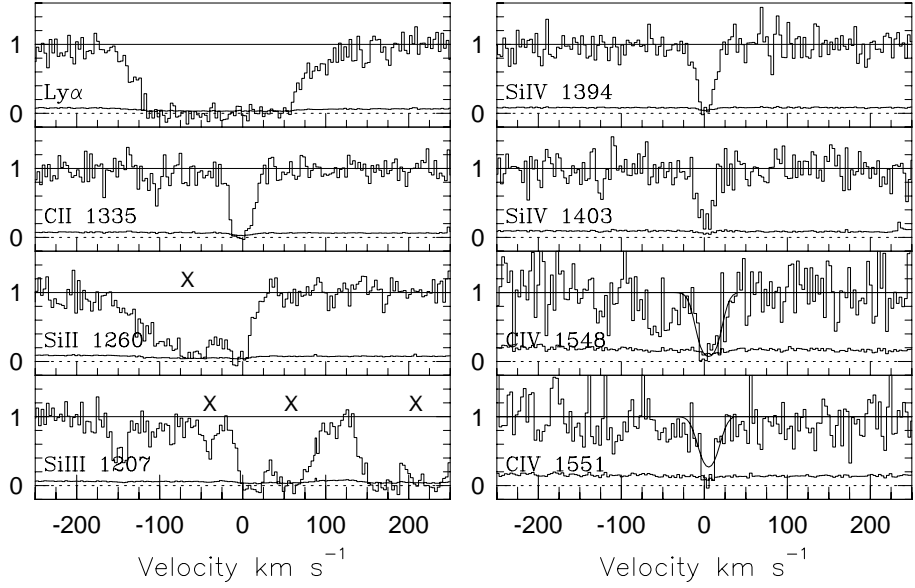}
\vglue -1in
\protect
\caption{Absorption profiles for the $z$ = 0.0809 system in the line of sight
  towards quasar PHL 1811, as in Figure~\ref{fig:f2}.  The {\SiIII}$\lambda$1207
  feature is blended to the red with Galactic {\SiII}$\lambda$1304 line.
Blends are marked with the symbol ``X''.}
\label{fig:f4}
\end{figure*}

\begin{figure*}
\figurenum{5}
\plotone{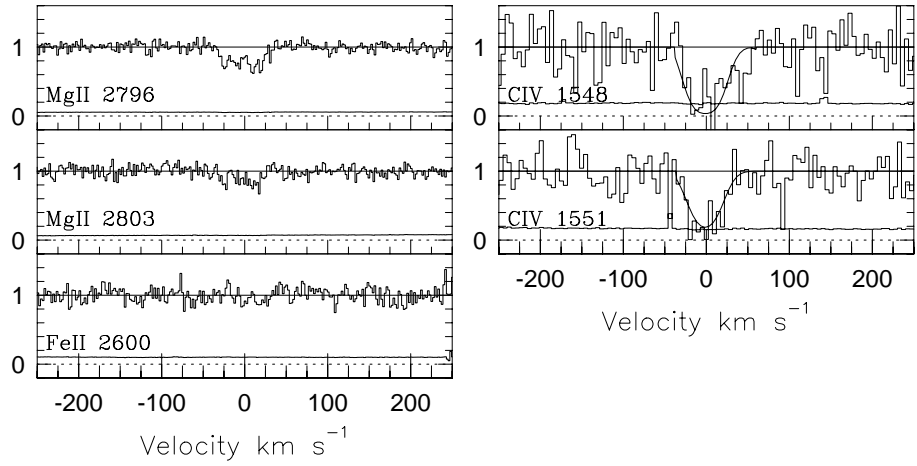}
\vglue -1in
\protect
\caption{The multiple-cloud weak low ionization system at $z$ = 0.5584
  in PG 1241+174 quasar spectra.  The low ionization transitions, {\FeII} 
  and {\MgIIdblt}, are covered by a Keck/HIRES spectrum, and other transitions
  by a {\it HST}/STIS spectrum.  Our Voigt profile fit to {\CIVdblt} is superimposed
  on those data.}
\label{fig:f5}
\end{figure*}

\begin{figure*}
\figurenum{6} \plotone{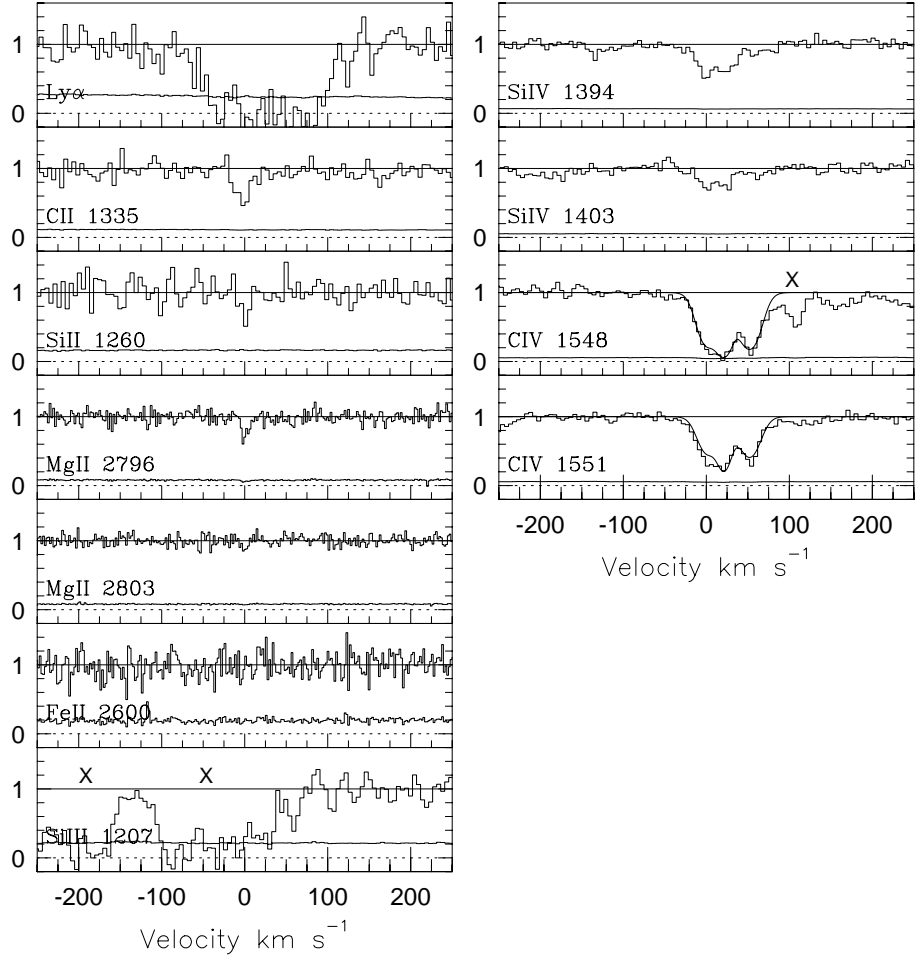} \vglue -1in \protect \caption{Same as
Figure \ref{fig:f5}, but for the system at $z$ = 0.6534 in spectrum
of PG 1634+706. Blends are marked with the symbol ``X''.} 
\label{fig:f6}
\end{figure*}

\begin{figure*}
\figurenum{7} \plotone{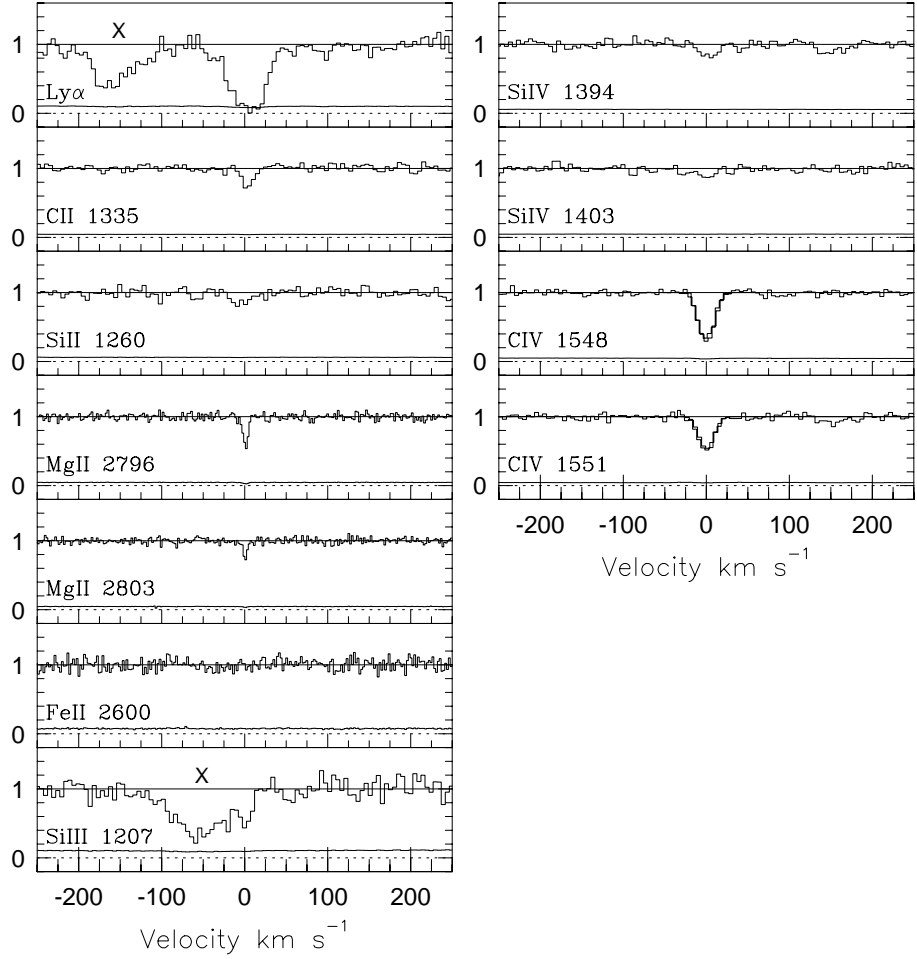} \vglue -1in \protect \caption{Same as
Figure \ref{fig:f5}, but the system $z$ = 0.8181 towards the quasar
PG 1634+706. Blends are marked with the symbol ``X''.} \label{fig:f7}
\end{figure*}

\begin{figure*}
\figurenum{8} \plotone{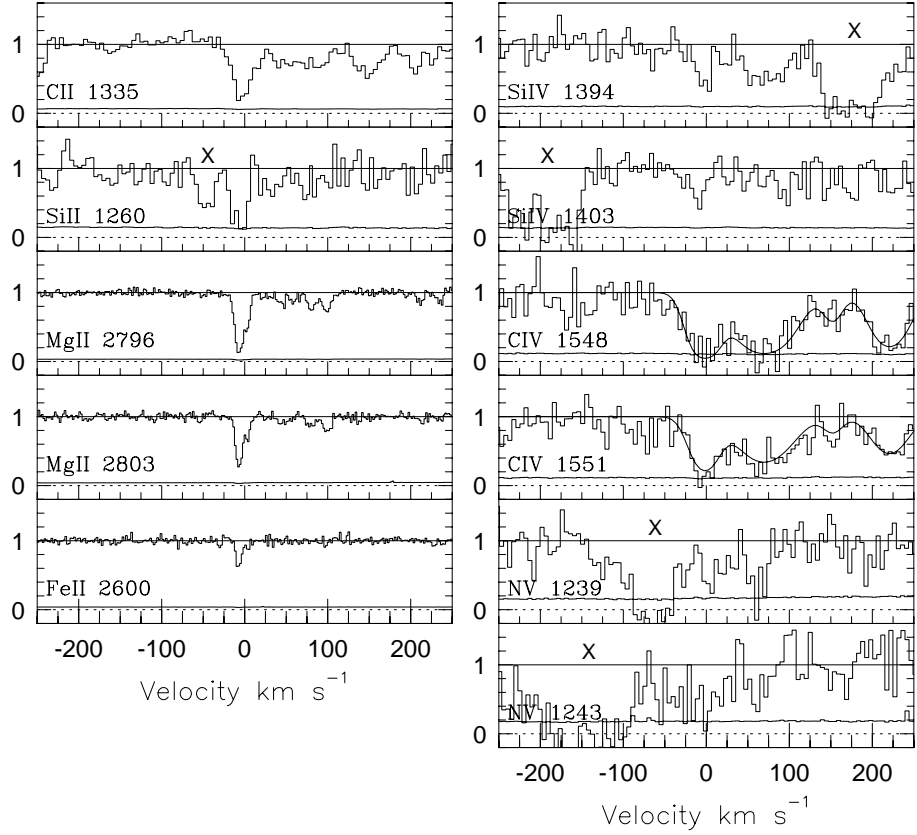} \vglue -1in \protect \caption{Same as
Figure \ref{fig:f5}, but for the system at $z$ = 0.8545 towards the
quasar PG 1248+401. Blends are marked with the symbol ``X''.} \label{fig:f8}
\end{figure*}

\begin{figure*}
\figurenum{9} \plotone{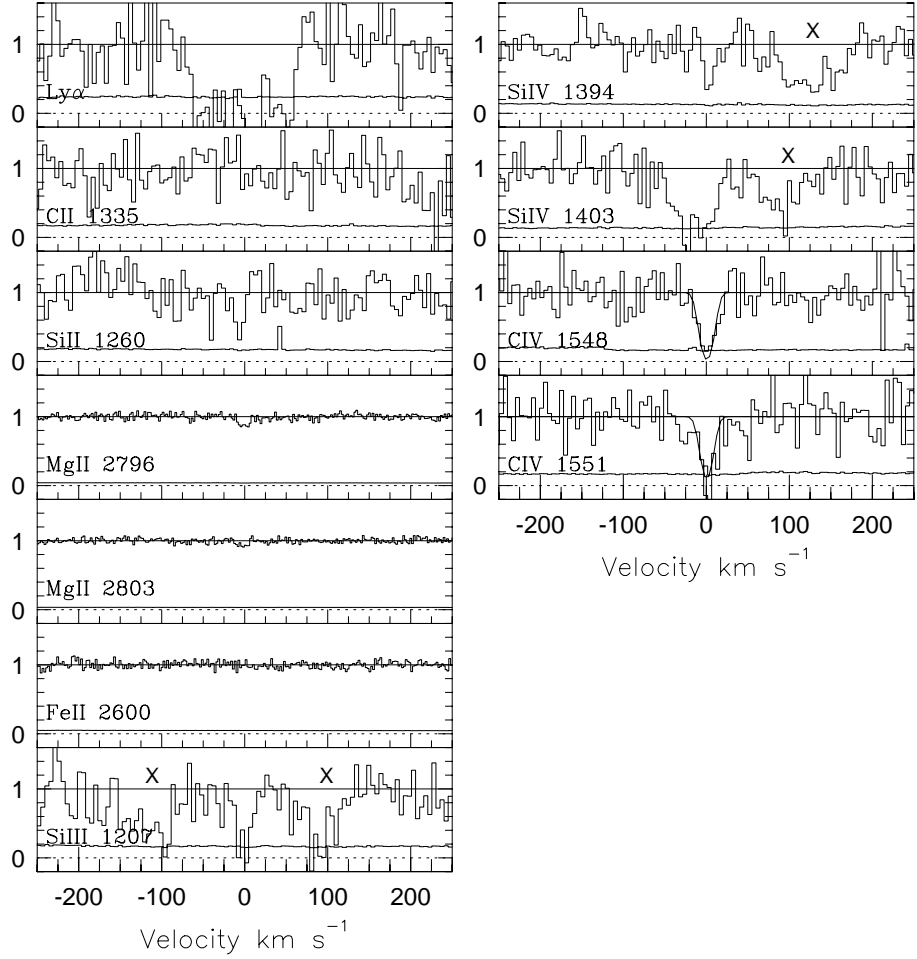} \vglue -1in \protect \caption{Same as
Figure \ref{fig:f5}, but for the system at $z$ = 0.8954 in the line
of sight of the quasar PG 1241+174. Blends are marked with the symbol ``X''.} \label{fig:f9}
\end{figure*}

\begin{figure*}
\figurenum{10} \plotone{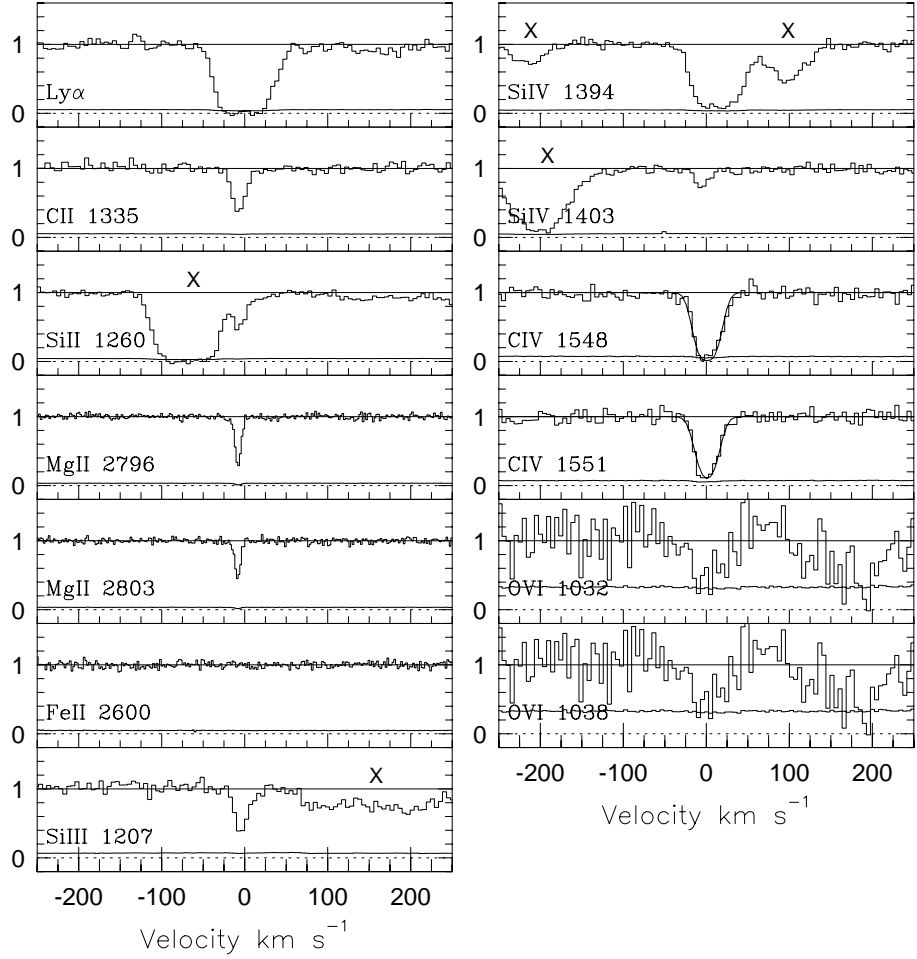} \vglue -1in \protect \caption{Same
as Figure \ref{fig:f5}, but for the system $z$ = 0.9056 towards the
quasar PG 1634+706. Blends are marked with the symbol ``X''.} \label{fig:f10}
\end{figure*}

\begin{figure*}
\figurenum{11} \plotone{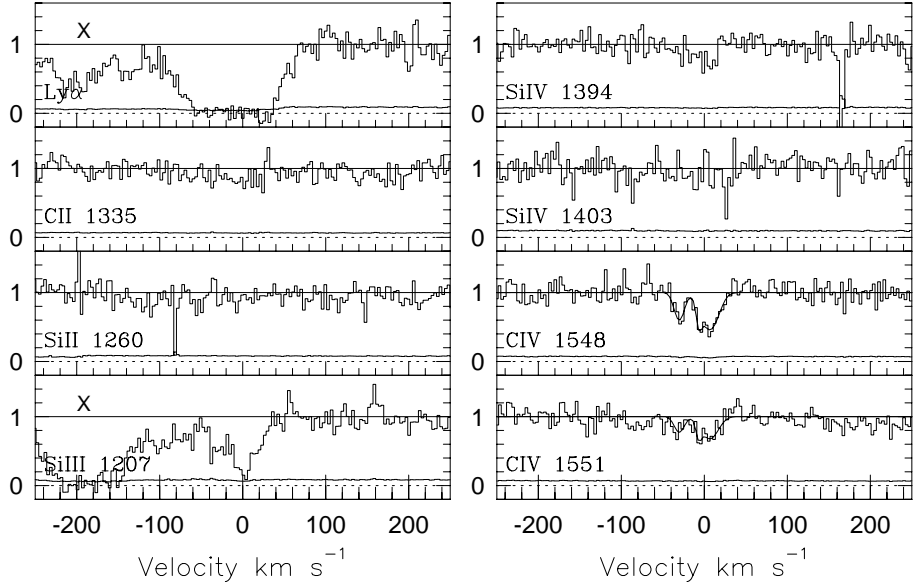} \vglue -1in \protect \caption{Same
as Figure \ref{fig:f2}, but for the system $z_{sys} = 0.0634$
towards the quasar HS 0624. Blends are marked with the symbol ``X''.} \label{fig:f11}
\end{figure*}

\begin{figure*}
\figurenum{12} \plotone{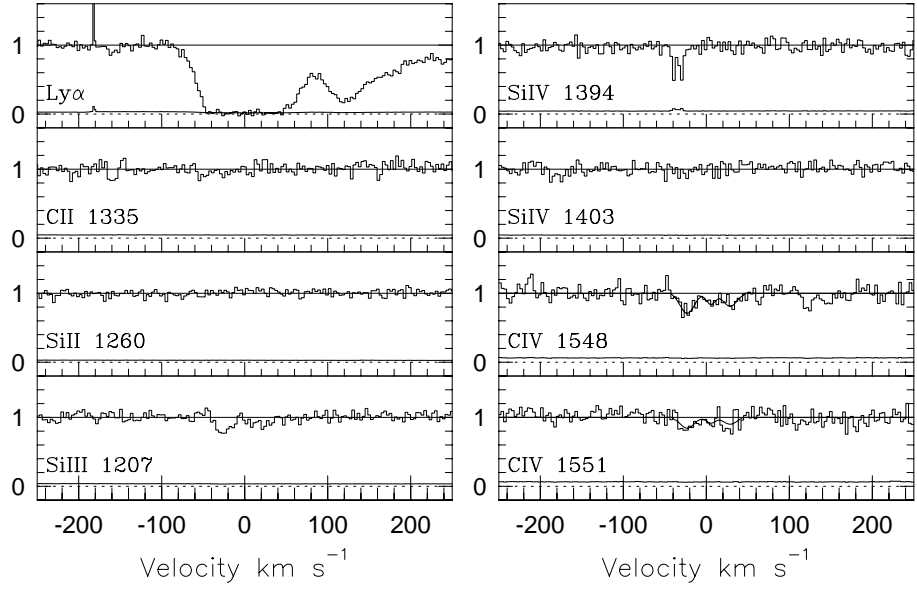} \vglue -1in \protect \caption{Same
as Figure \ref{fig:f2}, but for the multiple-cloud high only
ionization absorption system towards the quasar
  PG1211+143 at $z_{sys}=0.0644$.}
\label{fig:f12}
\end{figure*}

\begin{figure*}
\figurenum{13} \plotone{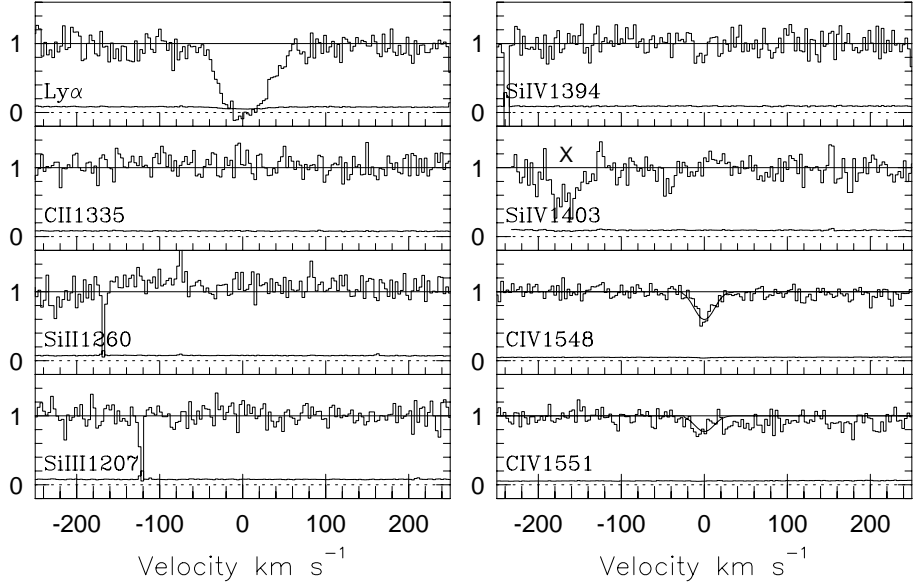} \vglue -1in \protect \caption{Same
as the Figure \ref{fig:f2}, but for the single-cloud high ionization
only system $z_{sys}=0.0757$ detected in
  the spectrum of quasar HS0624+6907. Blends are marked with the symbol ``X''.}
\label{fig:f13}
\end{figure*}

\begin{figure*}
\figurenum{14} \plotone{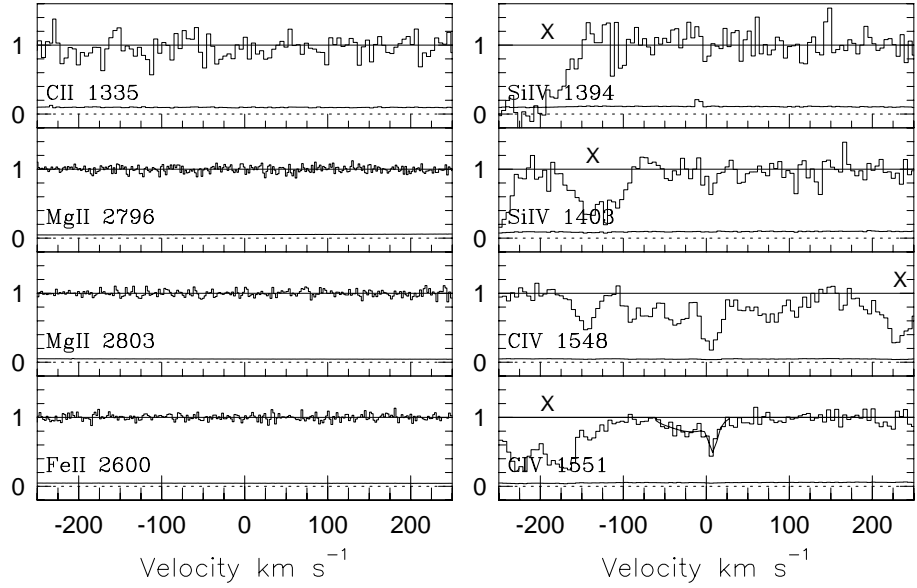} \vglue -1in \protect \caption{Same
as the Figure \ref{fig:f5}, but for the multiple-cloud high
ionization only system toward the quasar
  PG1206+459 at $z_{sys}=0.7338$. Blends are marked with the symbol ``X''.}
\label{fig:f14}
\end{figure*}

\begin{figure*}
\figurenum{15} \plotone{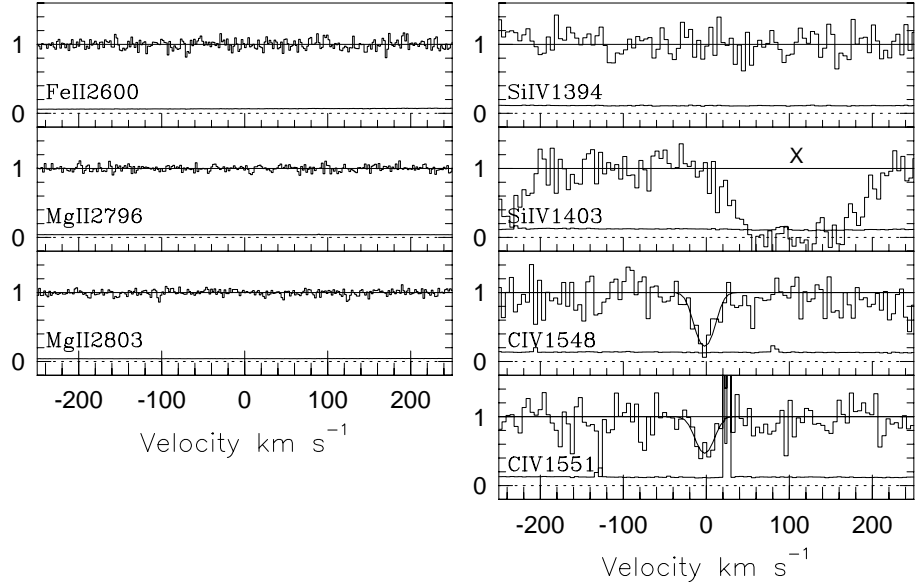} \vglue -1in \protect \caption{Same
as the Figure \ref{fig:f5}, but for the $z_{sys}=0.7011$
single-cloud high only system in the
  spectrum of the quasar PG1248+401. Blends are marked with the symbol ``X''.}
\label{fig:f15}
\end{figure*}

\begin{figure*}
\figurenum{16} \plotone{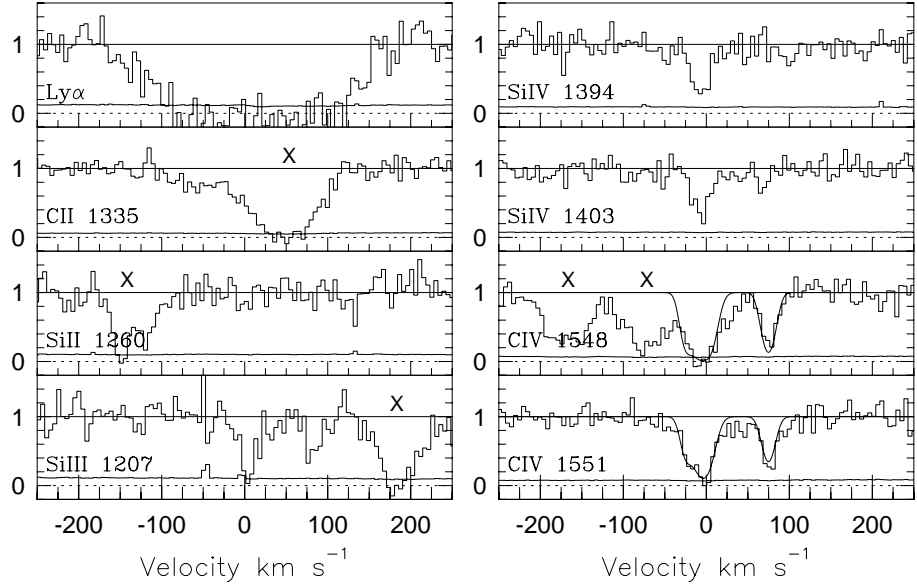} \vglue -1in \protect \caption{Same
as Figure \ref{fig:f2}, but for the system $z$ = 0.9143 towards the
quasar PG 1630+377. Blends are marked with the symbol ``X''.} \label{fig:f16}
\end{figure*}

\begin{figure*}
\figurenum{17} \plotone{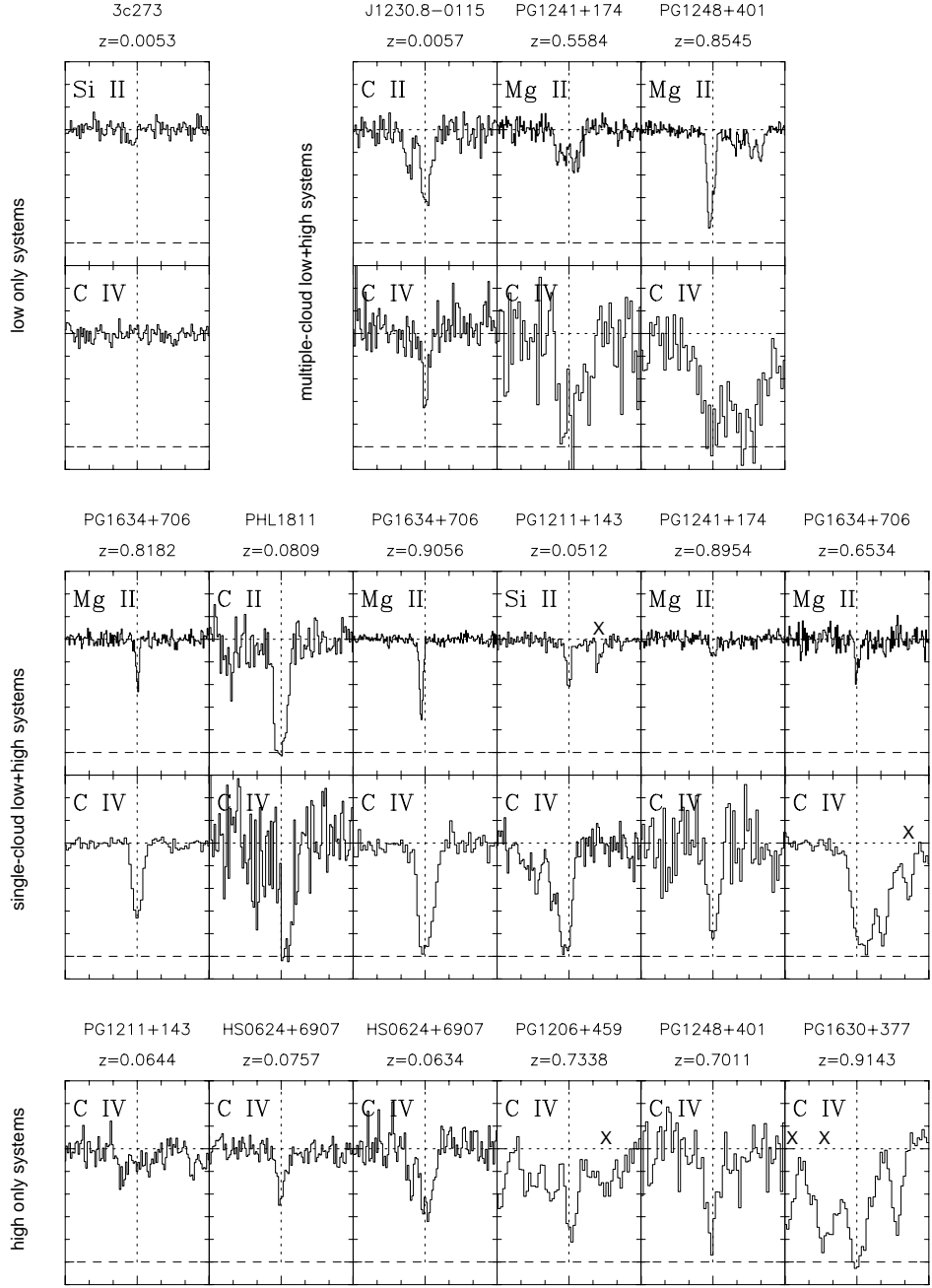} \vglue -1in \protect
\caption{The comparison of the kinematic profiles of {\CIV} between
the different classes of systems. The first row shows the low
ionization only system in the spectrum of the 3C 273 quasar, and the
multiple-cloud low + high absorption systems. The low ionization
transition with the highest quality data and the
{\CIV}$\lambda$1548 transition were used in this illustration.
Full system plots appear in Figures~\ref{fig:f1}--\ref{fig:f16}.
The second row presents the systems with single-cloud low + high
absorption, and the third row shows the {\CIV}$\lambda$1548
profiles for high only systems.The ``X'' marks indicate blends.} \label{fig:civkinplt}
\end{figure*}

\begin{figure*}
\figurenum{18} \plotone{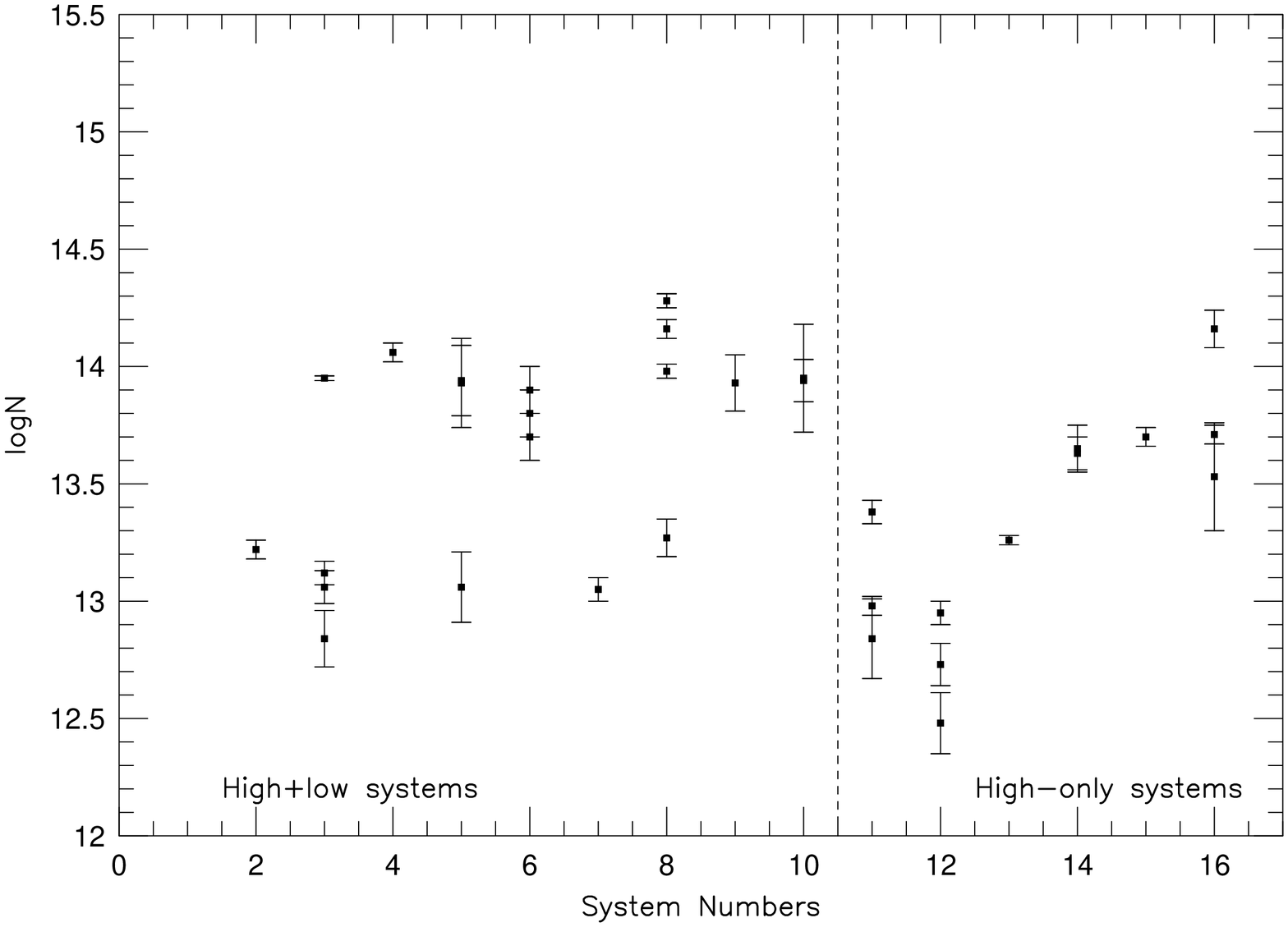} \protect \caption{The {\CIV}
column densities for the individual Voigt profile components in each
system. Systems are identified by number in Table~\ref{tab:vpfits}.
The comparison between the {\CIV} in low + high and high
only systems shows that high only systems tend to have component
{\CIV} column densities less than those of low + high ionization
systems.}

\label{fig:logn}
\end{figure*}

We also consider the relationship between the {\Lya} profiles and the
low ionization and {\CIV} profiles for our systems.  The {\Lya}
profiles are plotted along with those of the other transitions in
Figures~\ref{fig:f1}--\ref{fig:f16}.  The {\Lya} profiles encompass
the full velocity range of the metal line absorption.  In many cases,
the {\Lya} profiles are clearly not centered on the low ionization
absorption.  When there are offset {\CIV} components, the {\Lya}
profile is asymmetric in the same sense (see Figures~\ref{fig:f1} --
\ref{fig:f16}).  One example is the $z=0.9055$ system toward
PG~$1634+706$ (see Figure~\ref{fig:f10}), in which there is an offset
blended {\CIV} component to the red of the component centered on
{\MgII}.  Another example, in which the offset {\CIV} components are
distinct, is the $z=0.6534$ system toward PG~$1634+706$ (see
Figure~\ref{fig:f6}).

\begin{figure*}
\figurenum{19} \plotone{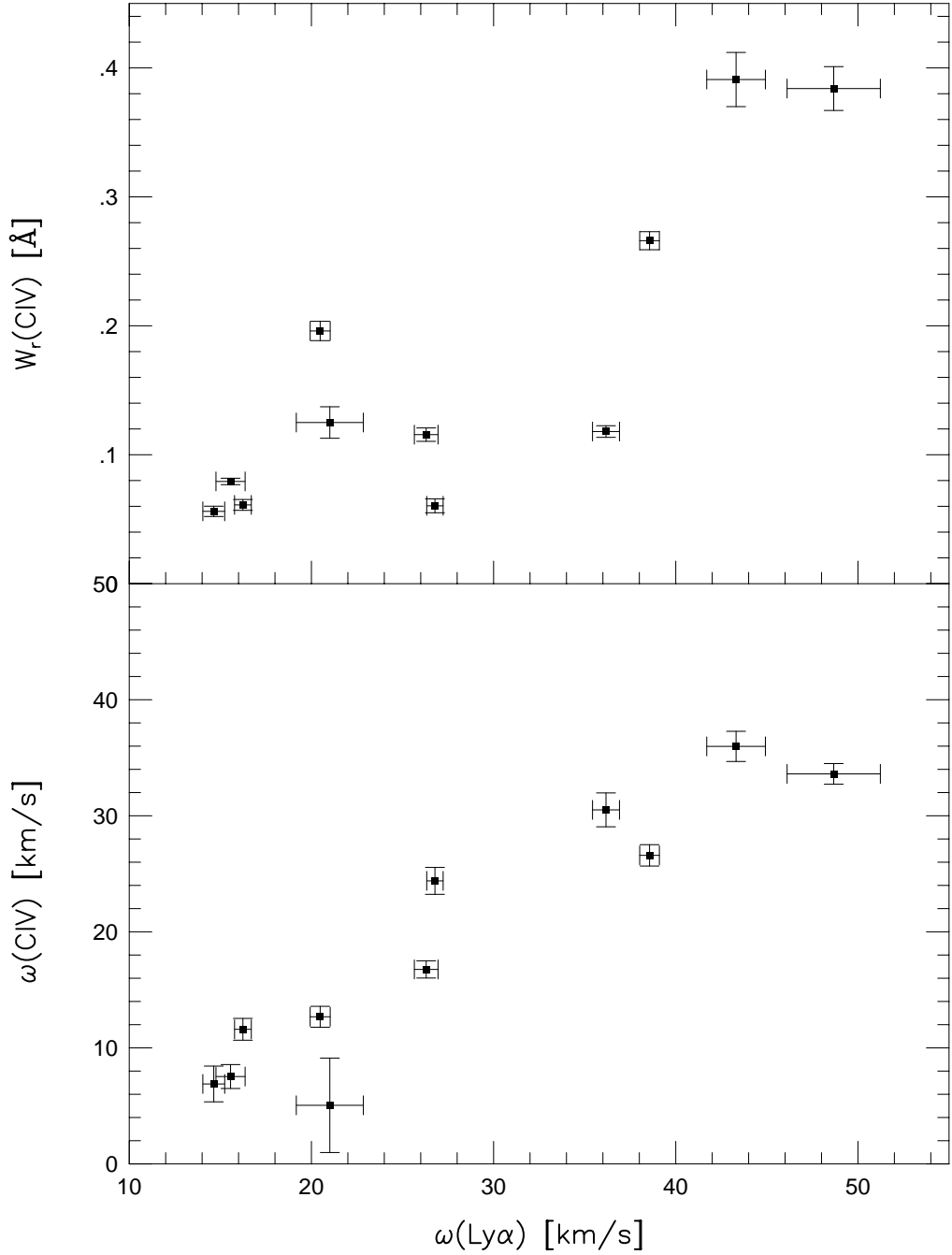} \protect

\vglue -0.5in
\caption{Top Panel-- Rest frame equivalent width of {\CIV}, $W_r(1548)$,
versus kinematic spread of {\Lya}, $\omega({\Lya})$.
Bottom Panel-- Kinematic spread of {\CIV}, $\omega({\CIV})$, versus kinematic
spread of {\Lya}, $\omega({\Lya})$.  This higher degree of correlation
between $\omega({\CIV})$ and $\omega({\Lya})$ implies that the kinematics
of the {\CIV} absorbing gas is a significant factor in determining the
strength of the {\Lya} absorption.}

\label{fig:kinlya}
\end{figure*}

In principle, the {\Lya} profile related to a given system might be
broader because the gas (either low or high ionization) has a lower
metallicity.  Alternatively, the {\Lya} profile might be broader
because of a large kinematic spread of the {\CIV} components, which
combine their associated {\Lya} absorption to produce the overall
{\Lya} profile.  To test between these two ideas, we consider whether
the kinematic spread of {\Lya} is more closely correlated with the
equivalent width of {\CIV} or with the kinematic spread of {\CIV}.
For fixed {\Lya}, we would expect $W_r({\CIV}\lambda{\rm 1548})$ to
be larger for a higher metallicity.  Figure~\ref{fig:kinlya} shows a
closer correlation between $\omega({\CIV}\lambda{\rm 1548})$ and
$\omega({\Lya})$.  To assess the significance of this apparent difference,
we conducted the Student's t-Test between two dependent correlations
\citep{chenpop}.  We found that, although both sets of values are
highly correlated, the correlation between $\omega({\CIV}\lambda{\rm
1548})$ and $\omega({\Lya})$ is more significant at the $3.0~\sigma$
level ($t=3.51$, yielding a probability of less than 1\%, for eight
degrees of freedom, that $\omega({\CIV}\lambda{\rm 1548})$ and
$W_r({\CIV}\lambda{\rm 1548})$ are more correlated).  The
kinematics of the {\CIV} absorbing gas appears to be a significant
factor in determining the strength of the {\Lya} absorption.

\section{Discussion of Geometry}
\label{sec:disc}

Considering the above results and the physical properties of weak low
ionization absorption systems, derived from photoionization models, we
will now discuss general constraints on their geometry.  Our goal is
to better understand the nature of the structures in which the low and
high ionization absorption arises.  Here we summarize briefly the
important observational constraints that must be satisfied:

\begin{enumerate}

\item{There are just six high only systems found in our survey sample, which
contains six single-cloud weak low ionization systems.  One additional system
is found with single-cloud weak low ionization absorption, but with no
detected high ionization absorption.  In comparison, three multiple
cloud weak low ionization systems were found, each with detected high
ionization absorption.}

\item{For low + high ionization systems, the strongest {\CIV}
absorption is aligned with the strongest low ionization absorption.}

\item{The {\CIV} profiles for both the high only and low + high
ionization systems have a similar distribution of kinematic spreads.
For both categories, some systems have single components of high
ionization absorption and some have multiple components.}

\item{The {\CIV} absorption tends to be similar, but somewhat weaker,
for high only systems than for low + high ionization systems, both in
the strongest {\CIV} component and in outlying {\CIV} components.}

\item{The {\Lya} absorption is very closely related to the kinematics
of the {\CIV} profiles, more so than it is to the low ionization
absorption.}

\end{enumerate}

Our discussion will focus on the origin of single-cloud weak low
ionization systems with properties adopted from \citet{weak2} and
\citet{weak1634}.  We choose this subset because many
multiple-cloud weak low ionization systems may have a different
physical origin, as described in \S~\ref{sec:intro}.  The structures
that produce single-cloud weak low ionization absorption were found to
have thicknesses ranging from $\sim 1$~pc -- $100$~pc, and
densities of $\sim0.01$~{\cc}.  The narrow line profiles imply a
velocity dispersion of $\sim 5$~{\kms}, which corresponds to a virial
mass of $\sim 3 \times 10^4$~{M$\odot$} for a $10$~pc structure.  This
is a considerably larger mass than the gas mass in a spherical
structure with density $0.01$~{\cc} and radius $10$~pc.  This suggests
confinement by dark matter mini-halos or by a stellar structure such
as a galaxy or star cluster.  The high ionization absorption related
to the same systems is produced in larger structures, with larger
velocity dispersions and with sizes of $\sim 1$~kpc, which could be
related to confinement.

In our survey, we found both low+high ionization systems and high only
systems.  Point 4 above implies that there are real differences, either
in degree or kind, between the low + high systems and the high only systems.
There appear to be two different kinds of {\CIV}
structures, those with low ionization regions of higher density covering
most of their area, and those without these higher density regions along the
line of sight.  
In this way we can explain the high only systems and
the weaker, offset components in low + high ionization systems as a
different or less extreme population than the central component of a
low + high ionization system.  In this population, eg., the total
hydrogen column density might have been lower so that a higher density
substructure did not collapse in the region.  The data would also be
consistent with a {\CIV} column density, for clustered clouds, that
tends to decrease outward from a central cloud in which {\MgII} absorption
arises.  Without such a fall-off, a model could not explain the
observed difference between the $N({\CIV})$ of the component centered
on the low ionization absorption and the offset {\CIV} and high only
system components.

Guided by these general principles, we now consider the most basic
``toy--model'' scenarios for the origins of the low and high
ionization absorption in single-cloud weak low ionization absorbers:

\begin{enumerate}

\item{{\it Single spherical low ionization region inside spherical
high ionization region}--- The first, simplest model to consider,
based on derived sizes, is one in which a $\sim 10$~pc low ionization
cloud is embedded in a $\sim 1$~kpc higher ionization halo.  This
model would give rise to $\sim 10^4\times$ more high only systems than
single-cloud low + high systems.  We observe a ratio of only $6/6$, such a
severe discrepancy that this model is easily ruled out.}

\item{{\it Multiple spherical low ionization regions inside spherical
high ionization region}---

The previous model can be improved by using multiple spherical low
ionization clouds inside a high ionization halo.  In order to produce
the derived ratio of high only to low + high ionization systems, the
covering factor should be $C_f \sim 0.5$.  This would imply a
probability of $\sim C_f^2 = 0.25$ of passing through two of the low
ionization clouds along a given line of sight.  This is not
inconsistent with the data, since some fraction of the $\sim 33$\% of
weak low ionization systems that have multiple clouds could be
produced in this way.  However, in order to produce $C_f \sim 0.5$,
there would need to be $\sim 5000$ low ionization structures with a
size of $\sim 10$~pc in each $\sim 1$~kpc high ionization halo.  If
the ratio of sizes of low to high ionization structures was instead
$1/1000$, there would have to be a factor of $100\times$ more small
clouds per large halo.  The average spacing between the small clouds
in the halo would be $94$~pc for the first choice of size ratio, and
$20$~pc for the second.  This is not much larger than the cloud size,
and seems a very contrived situation.  Without a physical model to
support such a scenario, we consider this model unlikely, though it is
not strictly ruled out.  Also, this simple model does not produce
offset high ionization components as are found in our survey.

We can modify this model to attempt to explain the observed offset
{\CIV} components.  In this modified model, separate high ionization
halos would give rise to these components.  These halos would have
lower covering factors for low ionization absorption than the ``main''
halo (the one giving rise to low ionization absorption).  Since we are
constrained to have an average of one offset component per low + high
system, roughly half of the sky, looking out from the main halo, would
need to be covered by the separate high ionization halos.  In this
case, looking from a large distance at the clustered group of halos,
an observer would see a covering factor for the separate high
ionization halos that greatly exceeds that from the main halo.  Adding
the separate high ionization halos would therefore produce more high
only systems than we observe, so that in fact we have not improved on
the model afterall.}

\item{{\it Low ionization shell surrounding high ionization shell}---

Alternatively, we now consider a situation where a low ionization shell
surrounds an interior shell of high ionization gas.  In order not to exceed the
number of multiple-cloud weak low ionization absorbers (relative to
single-cloud), the low ionization shell should be fragmented.  This
fragmented shell must cover roughly $33$\% of the surface area of the
bubble (assuming small fragments), so that it is more typical to pass
through low ionization absorption on only one side of the interior
region.

With only a single such ``bubble'', there is no clear explanation
for offset high ionization components for this model.
We might expect either one or two high ionization components, depending
on the covering factor of the high ionization shell.  It is not clear that
a component without detected low ionization absorption would have a
smaller {\CIV} column density than one which did, since it would come
from a different layer.
To account for the offset components, we can consider clustering of
``bubbles''.  However, such a model suffers from the same problems as
our similar attempts to modify Model 2, in that the additional bubbles
will produce too large a ratio of high only to low + high ionization
systems.}

\item{{\it Network of filaments/sheets which gives rise to high
ionization absorption, with embedded low ionization condensations}---

We finally consider a model in which low ionization condensations are
embedded in filaments and/or sheets which give rise to high ionization
absorption.  The overall covering factor of the low ionization regions
(relative to high ionization regions) must be $C_f \sim 0.5$ to
explain the observed ratio of high only to low + high ionization
systems.  In view of the small low ionization cloud thicknesses,
a filamentary/sheetlike geometry is a straightforward way to
produce the observed covering factor, without requiring huge
numbers of separate spherical clouds.
The lines of sight through filaments/sheets that do not give
rise to low ionization absorption are typically characterized by smaller
$N({\CIV})$ values than those that do.  These smaller $N({\CIV})$ structures
would give rise both to high only systems and to the offset high
ionization components of low + high systems.  In such a model, we
would expect the number of high ionization components to be similar
for the two categories of systems, as is observed.  We favor this model
because it naturally produces all of the results derived from our
survey.}

\end{enumerate}


\section{Connection to Physical Models}
\label{sec:disc2}

In the previous discussion, we intentionally avoided consideration of
the physical models that might give rise to weak low ionization
absorption.  Instead, we focused on basic geometric possibilities,
some of which we were able to exclude based upon results of our
survey.  Now, we consider the connections between these geometries and
the possible mechanisms for the origin of the absorbing clouds.

Based upon photoionization modeling, \citet{weak2} sketched out three
classes of physical models, for single-cloud weak low ionization
absorbers, that were consistent with the data.  In one class, the weak
{\MgII} absorption arose from the traces of gas remaining in a
Population III, pre-galactic or dwarf galaxy star cluster, with the
high ionization absorption coming from gas bound in the surrounding
dark matter halo.  In the second class of models, the weak {\MgII}
absorption is produced in fragments of shells from Type Ia supernovae,
housed in a surrounding dark matter halo.  In the third class of
models, the absorption arises in high velocity cloud structures that
are typically found in galaxy groups.

The most basic form of a star cluster model for weak {\MgII}
absorption matches Scenario 1 in \S~\ref{sec:disc}.  If there
is only one star cluster per spherical dark matter halo (which
produces the high ionization absorption) then this model is clearly
ruled out by cross-section arguments.  The idea of having many star
clusters per dark matter halo is described by Scenario 2.  It does not
seem reasonable to have such a tight packing of star clusters as would
be required.  Furthermore, this type of model cannot be consistent
with the small number of high only systems compared to low + high
ionization systems.  Although it agreed quite well with velocity
dispersions and derived cloud sizes of the low and high ionization
phases, it appears that star cluster models are ruled out.

The next model we consider is one in which supernova remnant shells
produce the weak, low ionization absorption, as described in Scenario
3 in \S~\ref{sec:disc}.  The {\CIV} absorption is most likely to be in
a concentric shell inside of the low ionization shell.  It is not
unreasonable that a supernova remnant would have a patchy shell
structure in order to produce the $33$\% covering of the surface area
as required by our survey results.  However, for this scenario, as
with Scenario 2, we were unable to envision a clustering of remnants
that would appropriately produce observed offset high ionization
absorption without overproducing high only systems.

Another physical picture to which Scenario 2 applies is
weak {\MgII} absorption arising in superwinds, as suggested by
\citet{stocke} and \citet{rosenberg} as an explanation for
absorption coincidence (low ionization and {\HI} at similar
velocities) between the $z=0.0057$ system toward RXJ~1230.8+0155 and
the nearby $z=0.0052$ system toward 3C~273, separated by $350
h^{-1}$~kpc.  Support for this hypothesis is provided by the
detection of a post-starburst galaxy at an impact parameter of $71
h^{-1}$~kpc from the 3C~273 line of sight. In that specific case, it
is hard to explain the absence of {\CIV} absorption along the 3C~273
line of sight in the context of a superwind cone/shell with a series
of ionized layers.  However, more generally, the superwind idea is
compelling since low ionization absorption is observed in nearby
superbubbles/winds \citep{heckman}. Clearly, such winds must present
a non-negligible cross section for absorption.  \citet{zonak} have
suggested that they may be responsible for some multiple-cloud weak
{\MgII} absorbers.  However, we argue here that there are problems
with using them to explain single-cloud weak low ionization
absorbers based upon our considerations of Scenario 2.  Again,
offset components require clustered additional bubbles/winds which
would produce too many high only systems.  This leads us to focus
primarily on Scenario 4.

There are similarities between recent models for high velocity cloud
structure in the Local Group \citep{fox,sembach} and Scenario 4.
\citet{fox} describes low ionization clouds sweeping through a hot
medium (either a galactic corona or an intergalactic medium) to
produce high ionization absorption in a conductive or turbulent
interface.  The Local Group high velocity clouds cover large areas of
the sky and exhibit similar absorption strengths and coherent motions
over large angular scales \citep{wakker, sembach, fox}, suggesting
sheetlike geometries.  We found that Scenario 4 was consistent with
our survey results.  \citet{weak2} stated that the single-cloud weak
low ionization absorbers with large {\FeII} column densities could not
be produced by high velocity cloud analogs.  This was because the high
iron abundance that was required implied ``in situ'' star formation
rather than ejection from Type II SNe.  Although present-day star
formation is limited in the Milky Way high velocity clouds, past star
formation is not ruled out, making a high velocity cloud model for
single-cloud weak low ionization absorbers more likely.  We also note
that {\OVI} absorption, which we have not examined in our survey,
should accompany {\CIV} absorption for analogs to Local Group high
velocity clouds.  Although Scenario 4 applies to high velocity cloud
models, it also applies more generally.  It remains to be understood how
high velocity cloud models fit in with the ``bigger picture'' of the
cosmic web of filaments and sheets expected from cosmological
simulations.

Another related possible scenario has weak low ionization absorption
arising in collapsed regions in material in tidal debris or as a
result of ram pressure due to high velocity infall through a hot
intracluster or intragroup medium.  Although it may not be common in
the present-day universe, there is at least one example of widespread
star formation in a group of galaxies moving through the rich cluster,
A1367 \citep{sakai}.  The star formation is taking place in many dwarf
galaxies and in intergalactic {\HII} regions, and it is of particular
interest that the metallicities in these star-forming regions are
close to the solar value \citep{sakai}, despite their dwarf
environments.  This implies that weak low ionization absorbers could
also arise in dwarfs, or at least in certain regions of some types of
dwarfs.  A faded population of such objects from similar processes
acting at high redshifts could present non-negligible cross-section.
The geometry of the remnant gas is uncertain, but could be consistent
with our Scenario 4.  This idea is not necessarily distinct from the
origin of high velocity clouds or from the general cosmic web.

The overlap between the single-cloud weak low ionization absorbers and
the population of {\CIV} absorbers at low redshift is substantial.
Thus information about the physical environments of {\CIV} absorbers
could provide clues to the nature of the single-cloud weak low
ionization absorbers.  By connecting absorption features with nearby
galaxies, \citet{chen} found that galaxies are surrounded by {\CIV}
halos with an extent of $R \sim 100 h^{-1} (L/L_B)^{0.5}$~kpc at
$W_r(1548) \sim 0.1$~{\AA}.  They favored a scenario in which these
{\CIV} halos were produced by gas stripped from accreting satellite
galaxies.  Also, it follows from their Figure~3 that all absorbers
with detected $W_r(1548)>0.1$~{\AA} were found within $R \sim 100
h^{-1} (L/L_B)^{0.5}$~kpc.  There are many additional galaxies at
larger impact parameters that have only $3~\sigma$ limits on
$W_r(1548)$, most of which fall between $0.1$-$0.2$~{\AA}.  However,
many of these may lie significantly below that detection limit.

In our sample, 5/6 of the low + high single-cloud weak low ionization
and 4/6 of the high only systems have $W(1548)>0.1$~{\AA}.  Based upon
the results of \citet{chen}, we would therefore expect that the
majority of these {\CIV} absorbers are produced within $R \sim 100
h^{-1} (L/L_B)^{0.5}$~kpc of a galaxy.  It also follows that a
significant fraction of the single-cloud weak low ionization absorbers
would be within this radius of a galaxy.  The fact that we did not
find a large number of high only systems implies that a large fraction
of the {\CIV} absorption regions around galaxies also produces single
cloud weak low ionization absorption.  This loose association between
weak low ionization absorption and luminous galaxies is consistent
with recent observations in individual cases \citep{churchill}, and is
not inconsistent with previous
claims that close associations are uncommon \citep{weak2}.

Our study indicates that, although weak
MgII absorbers are not directly associated with luminous galaxies,
they are in their environments.  
However, our statistics are relatively small, and more ultra-violet
spectra are needed to increase our understanding of the low redshift
weak low ionization absorber population.
Based on the geometrical
considerations in \S~\ref{sec:disc}, we favor a general model in which
spherical or flattened low ionization condensations or transient regions
inhabit sheets and/or filaments which give rise to high ionization.
In this model, the covering factor of the high ionization region by
low ionization condensations would need to be large, close to 50\%.
This type of origin in such a cosmic web is not necessarily different
from an origin in satellite dwarf galaxies or failed dwarf galaxies,
or in high velocity clouds, which are themselves components of the web.
Single-cloud weak low ionization
absorption requires condensations, apart from luminous galaxies, in
which stars have formed.  Such regions could be otherwise undetected
components of the cosmic web, yet to be resolved in numerical
simulations.  However, they could also be related to known structures
with uncertain origin, like the high velocity clouds, that are
already known to be clustered around galaxies.  In any case, we would
expect collapse and solar metallicity star formation products to be
most likely in the larger secondary potential wells that would exist in
the cosmic web in the vicinity of luminous galaxies.

\acknowledgements
Thanks to C. Churchill for providing his HIRES/Keck I
spectra.   We are also grateful to B. Jannuzi and S. Kirhakos for providing
fully reduced FOS spectra with continuum fits.  This work was enhanced by
our conversations with C. Churchill, M. \'Cirkovi\'c, A. Narayanan, and T. Tripp.
This research was funded by
NASA under grant NAG5-6399 NNG04GE73G and by NSF under grant AST 04-07138.
J.R.M. and R.S.L. were partially funded by the NSF REU program.



\clearpage

\clearpage

\clearpage

\clearpage

\clearpage

\clearpage

\clearpage

\clearpage

\clearpage

\clearpage

\clearpage

\clearpage

\clearpage

\clearpage

\clearpage

\clearpage

\clearpage

\clearpage

\clearpage

\begin{deluxetable}{llrrrlc}
\tablenum{1}
\tabletypesize{\footnotesize}
\tablewidth{0pt}

\tablecaption{{\it STIS}/HST E140M Grating Data}

\tablehead{
\colhead{QSO ID} &
\colhead{$z_{\textrm{\scriptsize QSO}}$} &
\colhead{S/N$^{\dag}$} &
\colhead{S/N$^{\dag}$} &
\colhead{S/N$^{\dag}$} &
\colhead{PI} &
\colhead{Proposal ID} \\
\colhead{} &
\colhead{} &
\colhead{1216~{\AA}} &
\colhead{1335~{\AA}} &
\colhead{1548~{\AA}} &
\colhead{} &
\colhead{}
}

\startdata
PKS 0405-123 & $0.534$ & $4.5$ & $6.4$ & $14.8$ & Heap & $7576$ \\
PG 0953+415 & $0.239$ & $5.9$ & $8.3$ & $5.8$ & Savage & $7747$ \\
PG 1116+215 & $0.177$ & $9.9$ & $10.9$ & $8.8$ & Sembach/Jenkins &$8097/8165$ \\
3C 273 & $0.158$ & $16.3$ & $24.4$ & $18.6$ & Heap &$8017$ \\
RX J1230.8+0115 & $0.117$ & $5.4$ & $9.5$ & $6.3$ & Rauch &$7737$ \\
PG 1259+593 & $0.478$ & $7.0$ & $9.0$ & $5.8$ & Tripp & $8695$ \\
PKS 1302-102 & $0.278$ & $5.2$ & $8.2$ & $8.1$ & Lemoine &$8306$ \\
3C 351.0 & $0.371$ & $5.0$ & $9.2$ & $6.5$ & Jenkins &$8015$ \\
H 1821+643 & $0.297$ & $8.1$ & $19.4$ & $11.5$ & Jenkins & $8165$ \\
PKS 2155-304 & $0.116$ & $13.2$ & $15.9$ & $14.1$ & Shull & $8125$ \\
PG 1211+143 & $0.081$ & $12.5$ & $24.8$ & $12.1$ & Shull & $8571$ \\
HS 0624+6907 & $0.370$ & $5.0$ & $8.8$ & $7.1$ & Tripp & $9184$ \\
3C 249.1 & $0.311$ & $5.6$ & $8.4$ & $5.2$ & Tripp & $4939$ \\
PG 1444+407 & $0.267$ & $4.1$ & $6.8$ & $7.9$ & Tripp & $9184$ \\
HE 0226-4110 & $0.495$ & $5.6$ & $7.7$ & $7.9$ & Tripp & $9184$ \\
PHL 1811 & $0.190$ & $6.4$ & $9.1$ & $7.1$ & Jenkins & $9418$ \\
PKS 0312-77 & $0.223$ & $4.5$ & $6.3$ & $4.2$ & Kobulnicky & $8312$ \\
TON S210 & $0.116$ & $4.0$ & $9.2$ & $5.0$ & Sembach & $9415$ \\
PG 1216+069 & $0.331$ & $4.2$ & $7.2$ & $4.6$ & Tripp & $9184$ \\
TON 0028 & $0.330$ & $4.4$ & $7.0$ & $4.2$ & Tripp & $9184$ \\
\hline
\enddata

\tablecomments{$\dag$ $S/N$ per pixel}
\label{tab:tab1}
\end{deluxetable}

\begin{deluxetable}{llcclr}
\tablenum{2}
\tabletypesize{\footnotesize}
\tablewidth{0pt}
\tablecaption{{\it STIS}/HST E230M Grating Data}

\tablehead{
\colhead{QSO ID} & \colhead{$z_{\textrm{\scriptsize QSO}}$} &
\colhead{S/N$^{\dag}$} & \colhead{S/N$^{\dag}$} & \colhead{PI} &
\colhead{Proposal ID} \\ \colhead{} & \colhead{} & \colhead{2382~\AA} &
\colhead{2796~\AA} & \colhead{} & \colhead{}
}

\startdata
PG 0117+210 & 1.491 & 8.5 & 14.2 & Jannuzi & 8673 \\
PKS 0232-04 & 1.450 & 6.1 & 8.7 & Jannuzi & 8673 \\
PKS 0312-77 & 0.223 & 3.1 & 5.4 & Kobulnicky & 8651 \\
PKS 0454-220 & 0.534 & \nodata & \nodata & Churchill & 8672 \\
HE 0515-4414 & 1.713 & 8.1 & 22.0 & Reimers & 8288 \\
HS 0747+4259 & 1.900 & 4.5 & 10.1 & Reimers & 9040 \\
HS 0810+2554 & 1.500 & 3.4 & 5.7 & Reimers & 9040 \\
PG 1116+215 & 0.117 & 10.1 & 15.1 & Sembach & 8097 \\
PG 1206+459 & 1.160 & 8.4 & 15.4 & Churchill & 8672 \\
PG 1241+176 & 1.273 & 5.2 & 20.0 & Churchill & 8672 \\
PG 1248+401 & 1.030 & 8.1 & 9.6 & Churchill & 8672 \\
CSO 873 & 1.022 & 8.0 & 10.4 & Churchill & 8672 \\
PG 1630+377 & 1.466 & 8.7 & 14.8 & Jannuzi & 8673 \\
PG 1634+706 & 1.334 & 19.8 & 28.5 & Jannuzi/Burles & 8312/7292 \\
PG 1718+481 & 1.083 & 15.9 & \nodata & Burles & 7292 \\
\hline
\enddata

\tablecomments{$\dag$ $S/N$ per pixel}
\label{tab:tab2}
\end{deluxetable}

\begin{deluxetable}{llrrrrrrr}
\tablenum{3}
\tabletypesize{\scriptsize}
\rotate
\tablewidth{0pt}
\tablecaption{Restframe Equivalent Widths of Selected Transitions}

\tablehead{
\colhead{\#} &
\colhead{QSO ID} &
\colhead{$z_{\textrm{\scriptsize system}}$} &
\colhead{${\Lya}$} &
\colhead{${\MgII}$} &
\colhead{${\SiII}$} &
\colhead{${\CII}$} &
\colhead{${\CIV}$} &
\colhead{${\SiIV}$}
}

\startdata

\multicolumn{8}{c}{Low Only Systems} \\
\hline
1 & 3c 273 & $ 0.0053 $ & $0.419 {\pm} 0.009$ & \nodata & \nodata &
$0.019 {\pm} 0.003$ & $ <0.0053 $ & \nodata \\
\hline
\multicolumn{8}{c}{Low + High Systems} \\
\hline
2 & RX J1230.8+0115 &$ 0.0057 $ & $ 0.615 {\pm} 0.006 $ & \nodata & $
0.103 {\pm} 0.008 $ & $ 0.105 {\pm} 0.005 $ & $ 0.056 {\pm}
0.004 $ & $ 0.078 {\pm} 0.005 $\\

3 & PG 1211+143 & $ 0.0512 $ & $ 1.230 {\pm} 0.005 $ & \nodata & $
0.048 {\pm} 0.003 $ & $ 0.157 {\pm} 0.003 $ & $ 0.266 {\pm}
0.007 $ & $ 0.083 {\pm} 0.004 $\\

4 & PHL 1811 & 0.0809 & $ 0.898 {\pm} 0.006 $ & &$ 0.158 {\pm} 0.007 $
&$ 0.164 {\pm} 0.005 $ &$ 0.118 {\pm} 0.005 $ &$ 0.119 {\pm}
0.002 $ \\

5 & PG 1241+176 & $ 0.5584 $ & \nodata & $ 0.132 {\pm} 0.006 $ & \nodata
& \nodata & $ 0.263 {\pm} 0.020 $ & \nodata \\

6 & PG 1634+706 & $ 0.6534 $ & $ 0.665 {\pm} 0.032 $ & $ 0.032 {\pm}
0.005 $ & $ 0.0207 {\pm} 0.008 $ & $ 0.045 {\pm} 0.006 $ & $0.384
{\pm} 0.017 $ & $ 0.123 {\pm} 0.008$ \\

7 & PG 1634+706 & $ 0.8181 $ & $ 0.216 {\pm} 0.010 $ & $ 0.036 {\pm}
0.001 $& $ 0.020 {\pm} 0.004 $ & $ 0.017 {\pm} 0.002 $ & $ 0.079
{\pm} 0.003 $ & $ 0.019 {\pm} 0.003 $ \\

8 & PG 1248+401 & $0.8545$ &  \nodata & $ 0.246 {\pm}
0.006 $ & $ 0.175 {\pm} 0.040 $ & $ 0.349 {\pm} 0.047 $ & $
0.873 {\pm} 0.024 $ & \nodata \\

9 & PG 1241+176 & $ 0.8954 $ & $0.547 {\pm} 0.027$ & $ 0.019 {\pm} 0.002 $ & $
0.038 {\pm} 0.007 $ & $ <0.021 $ & $ 0.125 {\pm} 0.012 $ & $
0.047 {\pm} 0.006$ \\

10 & PG 1634+706 & $ 0.9055 $ & $ 0.314 {\pm} 0.005 $ & $ 0.073 {\pm}
0.001 $ & $ 0.055 {\pm} 0.004 $ & $ 0.062 {\pm} 0.003 $ & $
0.196 {\pm} 0.008 $ & \nodata \\

\hline
\multicolumn{8}{c}{High Only Systems}\\
\hline

11 & HS 0624+6907 & $ 0.0635 $ & $0.582 {\pm} 0.022$ & \nodata & $ <0.006 $ & $<0.002 $ & $ 0.116 {\pm} 0.005$ & \nodata \\

12 & PG 1211+143 & $ 0.0644 $ & $0.590 {\pm} 0.012$ & \nodata & $<0.002$ & $<0.005$ & $ 0.060 {\pm} 0.006 $ & $ <0.013 $ \\

13 & HS 0624+6907 & $0.0757$ & $0.308 {\pm} 0.007$ & \nodata & $ <0.005 $ & $
<0.012 $ & $0.061 {\pm} 0.004$ & $0.048{\pm} 0.005$ \\

14 & PG 1206+459 & $ 0.7338 $ & \nodata & \nodata & \nodata & $< 0.011 $ &
$\sim0.25~\dag$ & $<0.021$ \\

15 & PG 1248+401 & $ 0.7011 $ & \nodata & $<0.012$ & \nodata & \nodata & $
0.114 {\pm} 0.009 $ & \nodata \\

16 & PG 1630+377 & $ 0.9143 $ & $ 1.119 {\pm} 0.019 $ & \nodata & $ 0.013 {\pm} 0.0096 $* & 
\nodata & $0.391 {\pm} 0.021$ & \nodata \\

\hline
\enddata
\vglue -0.05in
\tablecomments{* $3~\sigma$ level detection \\ 
               $\dag$ from the measurement of {\CIV}$\lambda$1550, due to the heavy blend in {\CIV}$\lambda$1548}

\label{tab:eqwidths}

\end{deluxetable}

\clearpage
\begin{deluxetable}{rllccrrr}
\tablenum{4}

\tabletypesize{\scriptsize} \tablewidth{0pt} \tablecaption{Voigt Profile Fits
of {\CIVdblt}}

\tablehead {
  \colhead{\#} & \colhead{QSO ID} & \colhead{z$_{\textrm{\scriptsize sys}}$} &
    \colhead{LogN} & \colhead{$\Delta$LogN} &\colhead{b} &
    \colhead{$\Delta$b} & \colhead{v} \\
               &                  &                                          &
    \colhead{[{\cmsq}]} & \colhead{[{\cmsq}]} & \colhead{[{\kms}]} &
    \colhead{[{\kms}]} & \colhead{[{\kms}]} \\
}

\startdata
\multicolumn{8}{c}{Low + High Systems}\\
\hline
2 & RX~J1230.8+0115 & 0.0057 & 13.22 & 0.04 & 6.87 & 0.91 & 0\\
\hline
3& PG~1211+143 & 0.0512 & 12.84 & 0.12 & 12.23 & 4.27 & -89 \\
& & & 13.06 & 0.07 & 9.56 & 1.76 & -65 \\
& & & 13.12 &0.05 & 6.84 & 1.03 & -36 \\
& & & 13.95 & 0.01 & 13.49 & 0.49 & -8\\
\hline
4 & PHL~1811 & 0.0809 & 14.06 & 0.04 & 11.24 & 0.76 & 0 \\
\hline
5& PG~1241+174 & 0.5584 & 13.93 & 0.19 & 7.97 & 2.32 & -10 \\
& & & 13.94 & 0.15 & 15.09 & 5.07 & 1 \\
& & & 13.06 & 0.15 & 7.09 & 3.6 & 25 \\
\hline
6 & PG~1634+706 & 0.6534 &$\sim$13.7  &   & $\sim$13 &    & -3 \\
  &             &        &$\sim$13.9  &   & $\sim$9  &    & 24 \\
  &         &        &$\sim$13.8  &   & $\sim$14 &    & 54 \\
\hline
7 & PG~1634+706 & 0.8181 & 13.05 & 0.05 & 6.60 & 1.80 & 0 \\
\hline
8& PG~1248+401 & 0.8545 &14.16 & 0.04 & 20.19 & 4.58 & -2 \\
 &             &        &14.28 & 0.03 & 38.86 & 3.03 & 70 \\
 &             &        &13.27 & 0.08 & 14.54 & 3.55 & 153 \\
 &             &        &13.98 & 0.03 & 26.59 & 1.74 & 221\\
\hline
9& PG~1241+174 & 0.8954 & 13.93 & 0.12 & 7.73 & 1.07 & 0 \\
\hline
10 & PG~1634+706 & 0.9056 & 13.95 & 0.23 & 5.50 & 1.80 & -2 \\
   & & & 13.94 &0.09 & 13.90 & 1.30 & 15 \\

\hline
\multicolumn{8}{c}{High Only Systems}\\
\hline
11& HS~0624+6907 & 0.0634 & 12.98 & 0.04 &  7.16 &  1.11 & -30 \\
& &                       & 12.84 & 0.17 &  1.79 &  2.04 & -6\\ 
& &                       & 13.38 & 0.05 & 12.99 &  1.55 & 7\\
\hline
12& PG~1211+143 & 0.0644 & 12.95 & 0.05 & 10.65 & 1.58 & -23 \\
  & & & 12.48 & 0.13 & 6.23 & 3.02 & 5\\
  & & & 12.73 & 0.09 & 10.45 & 3.00 & 29\\
\hline
13& HS~0624+6907 & 0.0757 & 13.26 & 0.02 & 15.06 & 0.91 & 0\\
\hline
14& PG~1206+459 & 0.7338 & 13.63 & 0.07 & 39.69 & 7.73 & -17 *\\
& & & 13.65 & 6.99 & 1.62 & 9.10 & 9 *\\
\hline
15& PG~1248+401 & 0.7011 & 13.70 & 0.04 & 13.26 & 1.45 & 0\\
\hline
16 &PG~1630+377 & 0.9143 & 13.53  &  0.23  &  6.59  & 3.44 &  -25 *\\
   &            &        & 14.16 &  0.08 &  12.52 &  2.17 &  -4 *\\
   &            &        & 13.71 &  0.04 &   8.42 &  0.70 &  75 \\

\enddata

\vglue -0.05in

\tablecomments{*Based on fit to {\CIV}$\lambda$1550 line since there
 were blends with {\CIV}$\lambda$1548}

\label{tab:vpfits}
\end{deluxetable}

\end{document}